\documentclass[twocolumn, dvipsnames]{aastex631}

\usepackage{multirow}
\usepackage{hyperref}
\usepackage{makecell}

\definecolor{DarkOrange}{RGB}{204, 85, 0}
\definecolor{LincolnGreen}{RGB}{17, 102, 0}
\definecolor{Rust}{HTML}{9B4F0F}
\definecolor{DarkCyan}{HTML}{008B8B}
\definecolor{MediumAquaMarine}{HTML}{66CDAA}
\definecolor{Maroon}{HTML}{800000}
\definecolor{Crimson}{HTML}{DC143C}

\newcommand{\BTSbot}{{\texttt{BTSbot}}}

\newcommand{\nfeatures}{25}

\newcommand{\btscompleteness}{$95.4\%$}
\newcommand{\btssavepurity}{$98.5\%$}
\newcommand{\btstriggerpurity}{$95.6\%$}

\newcommand{\testacc}{$94.9\%$}
\newcommand{\testrocauc}{$0.985$}
\newcommand{\testcompleteness}{100.0\%}
\newcommand{\ponetestpurity}{84.6\%}
\newcommand{\ptwotestpurity}{93.0\%}
\newcommand{\testmeddtsave}{-0.0381}
\newcommand{\testmeddttrigger}{-0.0147}

\newcommand{\pdayacc}{$96.2\%$}
\newcommand{\pdayrocauc}{$0.988$}
\newcommand{\pdaycompleteness}{$98.7\%$}
\newcommand{\ponepdaypurity}{$81.0\%$}
\newcommand{\ptwopdaypurity}{$92.0\%$}
\newcommand{\pdaymeddtsave}{$0.0525$}
\newcommand{\pdaymeddttrigger}{$1.01$}

\begin{document}

\title{The Zwicky Transient Facility Bright Transient Survey. III. \BTSbot: Automated Identification and Follow-up of Bright Transients with Deep Learning}

\author[0000-0002-5683-2389]{Nabeel Rehemtulla}
\affiliation{Department of Physics and Astronomy, Northwestern University, 2145 Sheridan Road, Evanston, IL 60208, USA}
\affiliation{Center for Interdisciplinary Exploration and Research in Astrophysics (CIERA), 1800 Sherman Ave., Evanston, IL 60201, USA}

\author[0000-0001-9515-478X]{Adam A. Miller}
\affiliation{Department of Physics and Astronomy, Northwestern University, 2145 Sheridan Road, Evanston, IL 60208, USA}
\affiliation{Center for Interdisciplinary Exploration and Research in Astrophysics (CIERA), 1800 Sherman Ave., Evanston, IL 60201, USA}

\author[0009-0003-6181-4526]{Theophile Jegou Du Laz}
\affiliation{Division of Physics, Mathematics, and Astronomy 249-17, California Institute of Technology, Pasadena, CA 91125, USA}

\author[0000-0002-8262-2924]{Michael W. Coughlin}
\affiliation{School of Physics and Astronomy, University of Minnesota, Minneapolis, Minnesota 55455, USA}

\author[0000-0002-4223-103X]{Christoffer Fremling}
\affiliation{Division of Physics, Mathematics, and Astronomy 249-17, California Institute of Technology, Pasadena, CA 91125, USA}
\affiliation{Caltech Optical Observatories, California Institute of Technology, Pasadena, CA 91125, USA}

\author[0000-0001-8472-1996]{Daniel A. Perley}
\affiliation{Astrophysics Research Institute, Liverpool John Moores University, IC2, Liverpool Science Park, 146 Brownlow Hill, Liverpool L3 5RF, UK}

\author[0000-0003-3658-6026]{Yu-Jing Qin}
\affiliation{Division of Physics, Mathematics, and Astronomy 249-17, California Institute of Technology, Pasadena, CA 91125, USA}

\author[0000-0003-1546-6615]{Jesper Sollerman}
\affiliation{Department of Astronomy, The Oskar Klein Center, Stockholm University, AlbaNova, SE-10691 Stockholm, Sweden}

\author[0000-0003-2242-0244]{Ashish~A.~Mahabal}
\affiliation{Division of Physics, Mathematics, and Astronomy 249-17, California Institute of Technology, Pasadena, CA 91125, USA}
\affiliation{Center for Data Driven Discovery, California Institute of Technology, Pasadena, CA 91125, USA}

\author[0000-0003-2451-5482]{Russ R. Laher}
\affiliation{IPAC, California Institute of Technology, 1200 E. California Blvd, Pasadena, CA 91125, USA}

\author[0000-0002-0387-370X]{Reed Riddle}
\affiliation{Caltech Optical Observatories, California Institute of Technology, Pasadena, CA 91125, USA}

\author[0000-0001-7648-4142]{Ben Rusholme}
\affiliation{IPAC, California Institute of Technology, 1200 E. California Blvd, Pasadena, CA 91125, USA}

\author[0000-0001-5390-8563]{Shrinivas R. Kulkarni}
\affiliation{Division of Physics, Mathematics, and Astronomy 249-17, California Institute of Technology, Pasadena, CA 91125, USA}



\begin{abstract}

The Bright Transient Survey (BTS) aims to obtain a classification spectrum for all bright ($m_\mathrm{peak}\,\leq\,18.5$\,mag) extragalactic transients found in the Zwicky Transient Facility (ZTF) public survey. BTS critically relies on visual inspection (``scanning'') to select targets for spectroscopic follow-up, which, while effective, has required a significant time investment over the past $\sim$5\,yr of ZTF operations. We present \texttt{BTSbot}, a multi-modal convolutional neural network, which provides a bright transient score to individual ZTF detections using their image data and \nfeatures~extracted features. \texttt{BTSbot}~is able to eliminate the need for daily human scanning by automatically identifying and requesting spectroscopic follow-up observations of new bright transient candidates. \texttt{BTSbot}~recovers all bright transients in our test split and performs on par with scanners in terms of identification speed (on average, $\sim$1 hour quicker than scanners). We also find that \BTSbot~is not significantly impacted by any data shift by comparing performance across a concealed test split and a sample of very recent BTS candidates. \texttt{BTSbot} has been integrated into Fritz and \texttt{Kowalski}, ZTF's first-party marshal and alert broker, and now sends automatic spectroscopic follow-up requests for the new transients it identifies. During the month of October 2023, \texttt{BTSbot} selected 296 sources in real-time, 93\% of which were real extragalactic transients. With \texttt{BTSbot} and other automation tools, the BTS workflow has produced the first fully automatic end-to-end discovery and classification of a transient, representing a significant reduction in the human-time needed to scan. Future development has tremendous potential for creating similar models to identify and request follow-up observations for specific types of transients. 

\end{abstract}



%% /-------- INTRODUCTION --------/ 
\section{Introduction} 
\label{sec:intro}

Large, wide-field surveys like Pan-STARRS \citep[Panoramic Survey Telescope and Rapid Response System;][]{Kaiser+2002}, ASAS-SN \citep[All-Sky Automated Survey for Supernovae;][]{Shappee+2014}, ATLAS \citep[Asteroid Terrestrial Last-Alert System;][]{Tonry+2011, Tonry+2018, Smith+2020},
% YSE \citep[Young Supernova Experiment;][]{Jones+2021}
and ZTF \citep[Zwicky Transient Facility;][]{Bellm+2019a, Bellm+2019b, Graham+2019, Dekany+2020, Masci+2019} have made immense contributions to time-domain astronomy by repeatedly imaging the entire night sky. Some surveys interface with the community through the output of an alert stream. Alert packets comprising the stream are intended to notify the community of the statistically significant brightening or dimming of some source with respect to a historical reference image. The alert stream provides time-series for a wide range of astrophysical, and non-astrophysical, phenomena. We are interested in the study of supernovae (SNe) using these alert streams, although only a small fraction of alerts in the unfiltered stream originate from genuine SNe. Thus, alert filters are deployed to identify candidate sources of interest. It is practically impossible to design an alert filter that rejects every irrelevant alert and accepts all alerts related to sources of interest. Therefore, manual candidate vetting, or ``scanning,'' of the filtered alerts is required. The term scanning can refer to different actions depending on the relevant survey and science case, but, in general, scanning has been used extensively and very successfully in the SN community to identify SNe for follow-up observations and further study.

The nightly data rate of the Vera C.~Rubin Observatory's Legacy Survey of Space and Time \citep[LSST;][]{Ivezic+2019} will significantly surpass our collective capacity for human scanning. Previous and ongoing surveys employ machine learning (ML) techniques for real/bogus classification \citep[e.g.,][]{Bailey+2007, Bloom+2012, Brink+2013, Wright+2015, Goldstein+2015, Cabrera-Vives+2017, Mahabal+2019, Duev+2019, Turpin+2020, Killestein+2021}, and adopting ML will be near-compulsory to efficiently extract knowledge from the next generation of surveys.

Beyond real/bogus classification, ML models have also been applied to a variety of tasks in astronomy including photometric transient classification \citep[e.g.,][]{Boone+2019, Muthukrishna+2019, Hosseinzadeh+2020, Villar+2019, Villar+2020, Gagliano+2023}, photometric redshift estimation \citep[e.g.,][]{CarrascoKind+2013, Sadeh+2016, Pasquet+2019}, and many others. Some have also been developed to, in part, reduce scanning load by encoding phenomenological or taxonomic information of a source in their model's output \citep[e.g.,][]{Bailey+2007, Duev+2021, Carrasco-Davis+2021, Gomez+2020, Gomez+2023, Stein+2023}.

For the most part, ML models in astronomy perform their tasks using extracted numeric features with architectures like random forests \citep{Breiman+2001}, or fully-connected neural networks \citep{McCulloch_Pitts_1943, LeCun+2015}. While appropriate in some cases, limiting these models to extracted features alone ignores potentially valuable information present in the images from which the features are extracted. A comparatively small number of convolutional neural networks \citep[CNNs;][]{Fukushima+1982} have been built that make use of the information embedded in astronomical images, and they have generally had great success \citep[e.g.,][]{Dieleman+2015, Lanusse+2018, Dominguez+2018, Duev+2019, Walmsley+2020}. CNNs are particularly well suited to astronomy because they can capture properties, like galaxy morphology, which often remain largely obscured to other image processing techniques \citep{Walmsley+2019}. Only a very small subset of these CNNs are \textit{multi-modal}, meaning they take in images and input of another type, like extracted features or a light curve \citep[e.g.,][]{van_Roestel+2021, Duev+2021, Carrasco-Davis+2021, Morgan+2022, Morgan+2023, Stoppa+2023}. We have developed a multi-modal CNN (MM-CNN) to automate scanning for a filtered stream from ZTF.

\begin{figure*}[ht]
    \begin{center}
    \includegraphics[width=2\columnwidth]{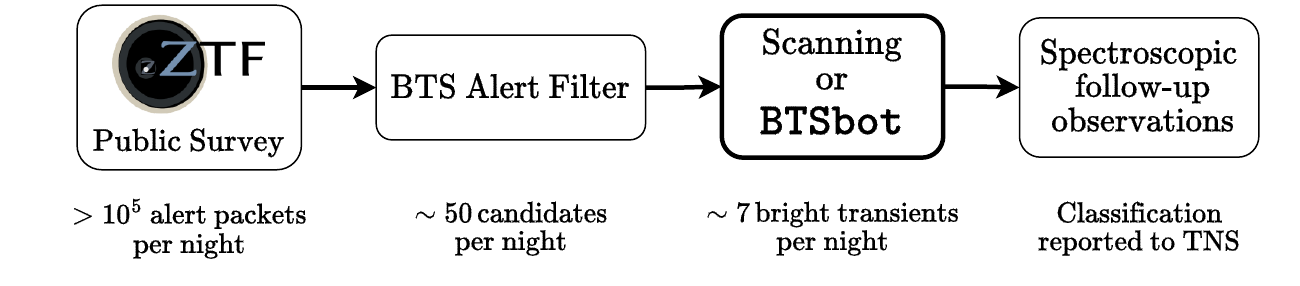}
    \caption{Diagram of the BTS workflow. The BTS alert filter ingests the ZTF public survey alert stream and, using upstream tools like \texttt{braai} and \texttt{sgscore}, removes bogus alerts, dim sources, and sources that are trivially not bright transients. Selection of real bright transients from the pool of candidates is done by visual inspection (``scanning"); in this study we develop an ML-based alternative: \BTSbot. The selected bright transients receive spectroscopic follow-up and classification, and are promptly reported to the public.}
    \label{fig:scanning}
    \end{center}
\end{figure*}

The Bright Transient Survey \citep[BTS;][]{Fremling+2020, Perley+2020} aims to spectroscopically classify all bright ($m_\mathrm{peak}\leq18.5\,\mathrm{mag}$ in $g$- or $r$-band) extragalactic transients\footnote{Henceforth, we use ``supernova" and ``transient" interchangeably with ``extragalactic transient." These are not equivalent, but it simplifies the prose.} from the ZTF public alert stream \citep{Patterson+2019}. The ZTF public survey produces $>10^5$ alert packets per night \citep{Mahabal+2019}. The majority of these are from variable sources or are bogus alerts, those arising from non-astrophysical phenomena. The BTS alert filter \citep{Perley+2020} removes from consideration most bogus alerts and many alerts from asteroids, active galactic nuclei (AGN), cataclysmic variables (CVs), variable stars (VarStars), and alerts from sources with $m>19.0\,\mathrm{mag}$ in $g$- and $r$-band. This filter makes use of \texttt{braai} \citep{Duev+2019} and \texttt{sgscore} \citep{Tachibana_Miller_2019} to filter the stream down to $\sim$50 new candidate BTS sources per night, of which $\sim$7 are new real bright transients. The other BTS candidates are mostly dim ($m_\mathrm{peak}>18.5\,\mathrm{mag}$) SNe or AGN, CVs, and VarStars that were not removed by the alert filter. All nightly BTS candidates are scanned by experts (``scanners'') who catalog (``save'') the real bright transients and request spectroscopic observations for classification. Scanning is performed on Fritz,\footnote{\url{https://github.com/fritz-marshal/fritz}} ZTF's first party marshal and a SkyPortal instance \citep{van_der_Walt+2019, Coughlin_2023}. BTS primarily executes its follow-up observations with robotic spectrographs like the SED machine \citep[SEDM;][]{Blagorodnova+2018, Rigault+2019, Kim+2022}, and, soon, the SED machine Kitt Peak\footnote{\url{https://sites.astro.caltech.edu/sedmkp/}} (SEDM-KP). The resulting classifications, whether assigned automatically by \texttt{SNIascore} \citep{Fremling+2021} or manually by a scanner, are promptly reported to the public via the Transient Name Server\footnote{\url{https://www.wis-tns.org}} (TNS), as visualized in Figure~\ref{fig:scanning}.

The BTS catalog is available online\footnote{\url{https://sites.astro.caltech.edu/ztf/bts}} and is updated in real-time. Since its origin in May 2018, BTS has maintained exceptionally high spectroscopic completeness of relevant sources (\btscompleteness\footnote{Computed for sources passing the BTS alert filter with $m_\mathrm{peak}\leq18.5\,\mathrm{mag}$, good light curve coverage, and displaying a SN-like light curve or a galaxy cross-match from May 2018 to September 2023. We adopt this value for the BTS sample completeness henceforth. See \cite{Perley+2020} for detailed definition of sample cuts.}), and BTS serves the community by rapidly releasing their classifications to the public. As of October 2023, the BTS sample contains more than 8,300 publicly classified SNe.

BTS enables a large amount of science, notably including some of the largest SN population studies conducted to date \citep[e.g.,][]{Perley+2020, Irani+2022,Sharon_Kushnir_2022,Sollerman+2022,Rodrigues+2023,Cold_Hjorth_2023,Sharma+2023}. The survey also provides unique discoveries \citep[e.g.,][]{Goobar+2023,Yang+2021} and is paving the way for using SNe to study large scale structure \citep{Tsaprazi+2022}.

Our new model, \BTSbot, enables the automation of scanning and spectroscopic follow-up for BTS by performing binary classification: bright transient~/~not bright transient. \BTSbot~produces a unit-interval bright transient score for an input ZTF Avro\footnote{\url{https://avro.apache.org}} alert packet \citep{Masci+2019, Patterson+2019} augmented with some custom metadata features. It is trained to run only on alerts from sources with at least one alert passing the BTS alert filter (candidate BTS sources), the same alerts which are provided to scanners. From image cutouts and extracted features, \BTSbot~must simultaneously separate transients from variable sources and, while $m>18.5\,\mathrm{mag}$, predict whether or not the source will attain $m_\mathrm{peak}\leq18.5\,\mathrm{mag}$. 

\BTSbot~is now integrated into \texttt{Kowalski} \citep{Duev+2019, Coughlin_2023}, ZTF's first-party alert broker, and Fritz, where \BTSbot~has now entered the BTS workflow. \BTSbot~is able to automatically save sources to BTS catalogs and conditionally trigger SEDM/SEDM-KP. \BTSbot~joins a rich collection of ML models and automation tools central to daily BTS operations, namely \texttt{braai} \citep{Duev+2019}, \texttt{sgscore} \citep{Tachibana_Miller_2019}, \texttt{pySEDM} \citep{Rigault+2019}, and \texttt{SNIascore} \citep{Fremling+2021}. Together, this workflow has yielded the first transient to be fully automatically detected, identified, spectroscopically classified, and publicly reported: SN~2023tyk \citep{Rehemtulla_AN23}. Zero human action was involved from the first detection to the publicly reported spectroscopic classification. Combining complementary ML models allows for the automation of a significant fraction of the tasks necessary to maintain BTS. This has the side-effect of freeing up time otherwise spent on repetitive, well-understood tasks, allowing for the better allocation of expert time and resources. We make the \BTSbot~source code and trained model publicly available on GitHub (\url{https://github.com/nabeelre/BTSbot}).

% The layout of this paper is as follows. Section~\ref{sec:train_data} describes how we compile a large and representative training set for \BTSbot~to learn from. Section~\ref{sec:architecture&training} details \BTSbot's architecture, the reasons why we chose such an architecture, and the techniques we employ to train \BTSbot. Section~\ref{sec:4} presents performance metrics for \BTSbot~on archival ZTF/BTS test split data (Sec.~\ref{sec:test_performance}) and a simulation of real-time performance with very recent present-day data (Sec.~\ref{sec:real_time_performance}). In Section~\ref{sec:5}, we discuss the integration of \BTSbot~into ZTF front- and back-end systems and the specifics around sending real triggers to spectrographs (Sec.~\ref{sec:integration}); other ML models similar to \BTSbot~in architecture (Sec.~\ref{sec:other_models}); and possibilities for adapting \BTSbot~to perform very rapid follow-up of young transients (Sec.~\ref{sec:2023ixf}). We summarize and conclude in Section~\ref{sec:summary}. 

%% /-------------------------------/
%% /-------- TRAINING DATA --------/
%% /-------------------------------/
\section{Training data} 
\label{sec:train_data}

The quality of an ML model's training set is a key factor in determining its performance. ML models, especially deep learning models like \BTSbot, are known to behave unpredictably when exposed to data unlike what they were trained on \citep{Szegedy+2013, Hendrycks+2016}. Thus, a model's training set must be fully representative of the model's input domain. \BTSbot's domain is ZTF alert packets from BTS candidates. These alerts come from SNe, AGN, CVs, VarStars, novae in very nearby systems, and a small number of other miscellaneous events including bogus alerts of many types (e.g., those due to poor image subtractions, high proper motion stars, saturated stars, etc.). We compile an extensive list of ZTF identifiers (ZTF-IDs) for sources that fall into these categories by drawing from a number of repositories. With these selections, we fully represent the wide variety of astrophysical phenomena \BTSbot~is exposed to in production.

An internal version of the BTS Sample Explorer is the most significant contributor to our initial list of ZTF-IDs. This internal BTS Sample Explorer operates identically to the public-facing version presented in Appendix~F of \citealt{Perley+2020}, but it adds information from internal ZTF catalogs. One of these catalogs is assembled by scanning an alert filter nearly identical to the BTS alert filter but altered to include sources with $19< m_\mathrm{peak}\,[\mathrm{mag}]<19.8$. Although these are not strictly BTS candidates, we allow them into the training set to increase the number of faint alerts. 

We draw from these catalogs with three queries. The first query, ``trues" for short, selects bright ($m_\mathrm{peak}\leq18.5\,\mathrm{mag}$) spectroscopically confirmed extragalactic transients that pass the purity cut (i.e. have a galaxy cross-match or a SN-like light curve, see Sec.~2.4 of \citealt{Perley+2020} for details). These bright confirmed transients make up the entire positive/true class of our training set. The second query, ``vars" for short, selects any source classified as an AGN, CV, or quasar (QSO) in any of the internal BTS catalogs. These sources are all considered non-transients or are not extragalactic and thus are part of the negative/false class regardless of their peak magnitude. The third query, ``dims" for short, selects any source that passes the purity cut with pre- and post-peak light curve coverage (see Sec.~2.3 of \citealt{Perley+2020} for coverage definitions) and $m_\mathrm{peak}>18.5\,\mathrm{mag}$. This is designed to broadly select dim SNe, which, because of their peak magnitudes, are part of the negative class. The \textit{dims} query is dominated by SNe but also includes a small number of other sources, e.g., tidal disruption events, CVs, asteroids, VarStars, etc. The requirement of pre- and post-peak light curve coverage is added to avoid selecting bright SNe with poor photometric coverage near peak, which could then appear as having dimmer peak magnitudes. Because peak magnitudes presented on the BTS sample explorer are computed from public data alone, there is a small population of sources which, due to alerts from partnership data, have $m_\mathrm{peak}\leq18.5\,\mathrm{mag}$ but appear in \textit{dims}. These are all removed entirely from the training set because many of them are unclassified.\footnote{These were identified after the test split was revealed so, to preserve the same splits, they are removed after train/validation/test splitting is performed.} By definition, neither \textit{vars} nor \textit{dims} overlap with \textit{trues}, but the imperfect nature of the selections done for \textit{vars} and \textit{dims} creates some overlap between the two. Sources in the overlap are removed from \textit{dims} and kept only in \textit{vars}. 
% This is not a source of label noise because any source from either of these queries is part of the negative class.

Any source queried from the BTS Sample Explorer was once marked as a bright transient by a human scanner. While we do build a sizeable list of non-bright~transients saved by the human scanners, there is a missing population of BTS candidates that are never saved. Our fourth query, ``rejects" for short, represents these sources. This list is limited to sources with alerts that pass the BTS alert filter between January~1st~2021 and January~1st~2023 UTC\footnote{All dates and times here are in UTC unless otherwise specified.}. Many candidates from before January~1st~2021 encountered earlier versions of the BTS alert filter, which was last improved in late 2020; many candidates from after January~1st~2023 were still evolving at the time the sources were queried and thus have potentially uncertain types. By not saving them, scanners implicitly mark these sources as non-bright~transients and thus part of our negative class. This assumption is reasonable given BTS's very high photometric completeness. An extremely small fraction of the sources in the \textit{rejects} list may be bright transients and inject a small amount of label noise into our training set, but we find that this effect, if present at all, is small enough to be ignored.

There is an additional population of BTS candidates, which we do not include in our training set: ``junk'' for short. These sources, which frequently pass the BTS alert filter, have been identified by scanners to clearly not be bright transients. They are mostly AGN and VarStars but also include smaller numbers of sources like high-proper motion stars and long-lived CVs. Experiments trained on these sources frequently yielded worse performance than equivalent models which excluded them. Some sources selected by our other queries are also cataloged in \textit{junk}; these sources remain in the other query and are not removed with other \textit{junk} sources. 
% In practice, BTS scanners typically filter out sources cataloged as \textit{junk} and do not consider them when scanning. We emulate this by ignoring sources cataloged in \textit{junk} during calculations of sample purity (see Sec.~\ref{sec:test_performance}, \ref{sec:real_time_performance}).

% The final and smallest contributor to our training set is a list of externally identified Type-Ia supernovae, ``extIas" for short. These sources were initially found in other surveys (like ?? \textbf{REF}) and were later identified in ZTF data by cross-matching using their sky position and time of peak. \nr{more info and refs needed here from Mat Smith}. Because they are all known to be SNe, we include them in the positive class if they have a ZTF detection with $m\,<\,18.5\,\mathrm{mag}$, and we include them in the negative class otherwise. \nr{would be slightly better to label them with peakmags queried from TNS? or another survey? probably more work than its worth}

Once we have a list of ZTF-IDs and their corresponding labels, we query \texttt{Kowalski} to retrieve all alert packets from each source. BTS only uses data from the ZTF public survey, but we include data from partnership time to increase the size of our training set. We arrange the science, reference, and difference image cutouts of each alert packet to form a $63\times63\times3$ image, or a triplet. Image cutouts are individually normalized with Euclidean-normalization\footnote{Also called $L^2$-normalization.}; pixels in masked regions are uniformly given values of 0; cutouts smaller than $63\times63$ (due to being near the edge of a CCD) are padded to $63\times63$ with pixels of value $10^{-9}$. This image pre-processing is identical to \texttt{braai}'s \citep{Duev+2019}. Some cutouts have pixel values uniformly set to \texttt{NaN} or exactly zero. These cutouts are corrupted, and alerts with any corrupted cutouts are removed. The distribution of alerts and sources in the training set at this stage is shown in Table~\ref{tab:train_data} under the initial queries header. 

\begin{deluxetable}{ccc}
\tablecaption{Training set size before/after cleaning cuts \label{tab:train_data}}
\tablehead{
    \colhead{Name of Query} & \colhead{Number of Sources} & \colhead{Number of Alerts}
}
\startdata
    \multicolumn{3}{c}{Initial queries} \\
    \hline
    \textit{trues}\tablenotemark{a} & 5,212 & 308,934 \\
    \textit{vars}\tablenotemark{b} & 1,127 & 150,017 \\
    \textit{dims}\tablenotemark{c} & 8,979 & 249,087 \\
    \textit{rejects}\tablenotemark{d} & 4,417 & 407,357 \\
    % \hline
    Total & 19,735 & 1,115,395 \\
    \hline
    % \hline
    \multicolumn{3}{c}{Cleaned training set} \\
    \hline
    \textit{trues}\tablenotemark{a} & 5,206 & 264,317 \\
    \textit{vars}\tablenotemark{b} & 1,126 & 109,934 \\
    \textit{dims}\tablenotemark{c} & 8,824 & 223,934 \\
    \textit{rejects}\tablenotemark{d} & 4,402 & 241,478 \\
    % \hline
    Total & 19,558 & 839,663 \\
    % \hline
\enddata
\tablenotetext{a}{Spectroscopically confirmed bright ($m_\mathrm{peak}\,\leq\,18.5\,\mathrm{mag}$) extragalactic transients.}
\tablenotetext{b}{Sources classified as AGN, CVs, VarStars, or QSOs.}
\tablenotetext{c}{Dim ($m_\mathrm{peak}\,>\,18.5\,\mathrm{mag}$) sources with transient-like light curves.}
\tablenotetext{d}{Sources not marked as bright extragalactic transients by BTS scanners.}
\tablecomments{Alerts are removed from the training set if they (i) have a corrupted image cutout; (ii) come from a source with an ambiguous label; (iii) are missing Pan-STARRS1 cross-match information; (iv) are an $i$-band observation; (v) have a negative difference image; or (vi) come from a source with a transient present in the reference image.}
\end{deluxetable}

\begin{figure}[ht]
    \begin{center}
    \includegraphics[width=0.9\columnwidth]{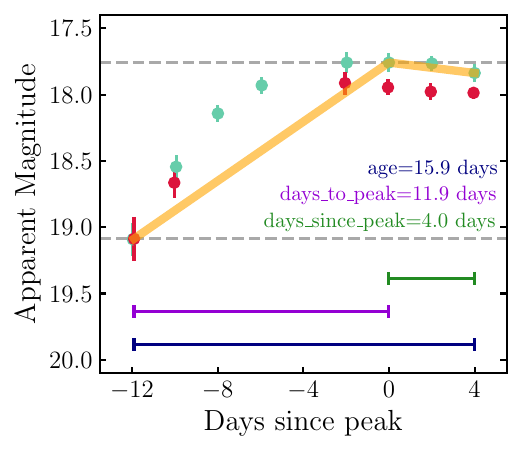}
    \caption{Custom metadata feature definitions depicted for the light curve of ZTF20acjlkpe. Teal ($g$-band) and red ($r$-band) circles indicate detections. \texttt{days\_to\_peak} (purple), \texttt{days\_since\_peak} (green), \texttt{age} (navy), \texttt{peakmag\_so\_far} (upper dashed gray), and \texttt{maxmag\_so\_far} (lower dashed gray) are presented for the latest detection shown. Together, these features make simplified information of the light curve phase and shape available to \BTSbot~(shown as the orange line, which assumes that \texttt{maxmag\_so\_far} corresponds to the first detection).}
    \label{fig:metadata_LC}
    \end{center}
\end{figure}

We then augment the alert packets with custom metadata features that are not already present in the ZTF Avro alert packets: \texttt{days\_to\_peak}, \texttt{days\_since\_peak}, \texttt{age}, \texttt{peakmag\_so\_far}, \texttt{maxmag\_so\_far}, and \texttt{nnondet}. The feature \texttt{days\_to\_peak} encodes the number of days from the first alert to the alert with the brightest magnitude thus far for the source in question. Closely related is \texttt{days\_since\_peak}, which represents the number of days from the current alert to the alert with the brightest magnitude. The sum of these quantities is the \texttt{age}. The features \texttt{peakmag\_so\_far} and \texttt{maxmag\_so\_far} encode the brightest and dimmest magnitude of all alert packets from this source thus far; \texttt{peakmag\_so\_far} is particularly crucial for correctly classifying late-time alerts of SNe because \BTSbot~would otherwise have almost no information on the brightness history of the source. These features are shown over an example light curve in Figure~\ref{fig:metadata_LC}. They clearly do not recover the light curve perfectly, but they are adopted because of their simplicity and effectiveness across a very large variety of light curves. We find that giving \BTSbot~high-level information on the source's current phase and the rough shape of its light curve with these features yields a significant boost in performance. Many methods exist to more faithfully represent SNe light curves, but many are not conditioned to model light curves of the other sources in \BTSbot's domain, e.g. AGN, VarStars, and others. Notably, these features are defined for all sources with as few as one detection or detections in only one photometric passband. One could extract more information from the light curve by decoupling these features to passband-dependent equivalents, however, this would create instances of alerts having missing features potentially compromising the model's performance. Lastly, the feature \texttt{nnondet} is an estimation of the number of non-detections at the source's location. This feature was used in the ALeRCE real-time stamp classifier \citep{Carrasco-Davis+2021} and was found to have high importance for distinguishing alerts between their five classes; we find that including it similarly boosts \BTSbot's performance.

All of these features are computed for each alert packet from its perspective, i.e., only using information already available at the time the alert packet was created, including alerts from before each source passed the BTS alert filter.
% public and partnership? or just public?
Additionally, some alerts were created before the latest version of \texttt{braai} went into production and thus have a deep real-bogus score (\texttt{drb}) from an outdated version of \texttt{braai} or no \texttt{drb} at all. We rerun all alert packets through the \texttt{d6\_m9} version of \texttt{braai} and replace all \texttt{drb} scores with these new scores.

Next, we clean our training set with a series of cuts. First, we remove alerts from $i$-band observations and alerts with negative difference images. The $i$-band alerts are removed primarily because BTS draws from the ZTF public survey, which only takes observations in $g$- and $r$-band. Negative difference images are identified using the \texttt{isdiffpos} flag included in the ZTF alert packets. Alerts with negative difference images are removed because they frequently represent instances of SNe or other transients visible in the reference image: a configuration different than what is typical for an alert. As an extra precaution, we compile a list of sources with SNe present in the reference images and remove them from the training set. Additionally, some sources fit into neither or both of our classes and are correspondingly removed from our training set. For example, ZTF18abdiasx appears to be a bright supernova projected on top of an AGN, and thus fits in both of our classes so it is removed from the training set. Some of the metadata features we choose to include reference the Pan-STARRS1 catalog \citep[PS1;][]{Kaiser+2002}: \texttt{sgscore\{1,2\}} and \texttt{distpsnr\{1,2\}}. When there are no PS1 sources within 30 arcseconds of a given ZTF alert, these features are set to $-999$. We remove all alerts with $-999$ in any of these four fields. Table~\ref{tab:train_data} shows the counts of sources and alerts at this stage under the cleaned training set header. Each of the four queries loses some alerts in the cleaning process, however, the \textit{rejects} query loses more than $150,000$ alerts. These $150,000$ alerts are almost entirely negative difference image alerts, which come from sources like binary stars, AGN, and high-proper motion stars. 
%After cleaning, our queries yield 19,701 sources (5,206 BTS and 14,495 not-BTS) and 848,483 alerts (264,316 BTS and 584,167 not-BTS).

We employ standard practices for splitting the full training set into train, validation, and test splits. We randomly assign sources to these splits with probabilities of 81\%, 9\%, and 10\% respectively\footnote{90\% of the training data is seen by the model during development and the remaining 10\% is concealed until hyperparameters are finalized.}. This splitting is done on sources rather than alerts to prevent any source from having alerts in multiple splits, which would produce a validation and test bias that overestimates the true performance of the model.

\begin{figure}
    \begin{center}
    \includegraphics[width=0.9\columnwidth]{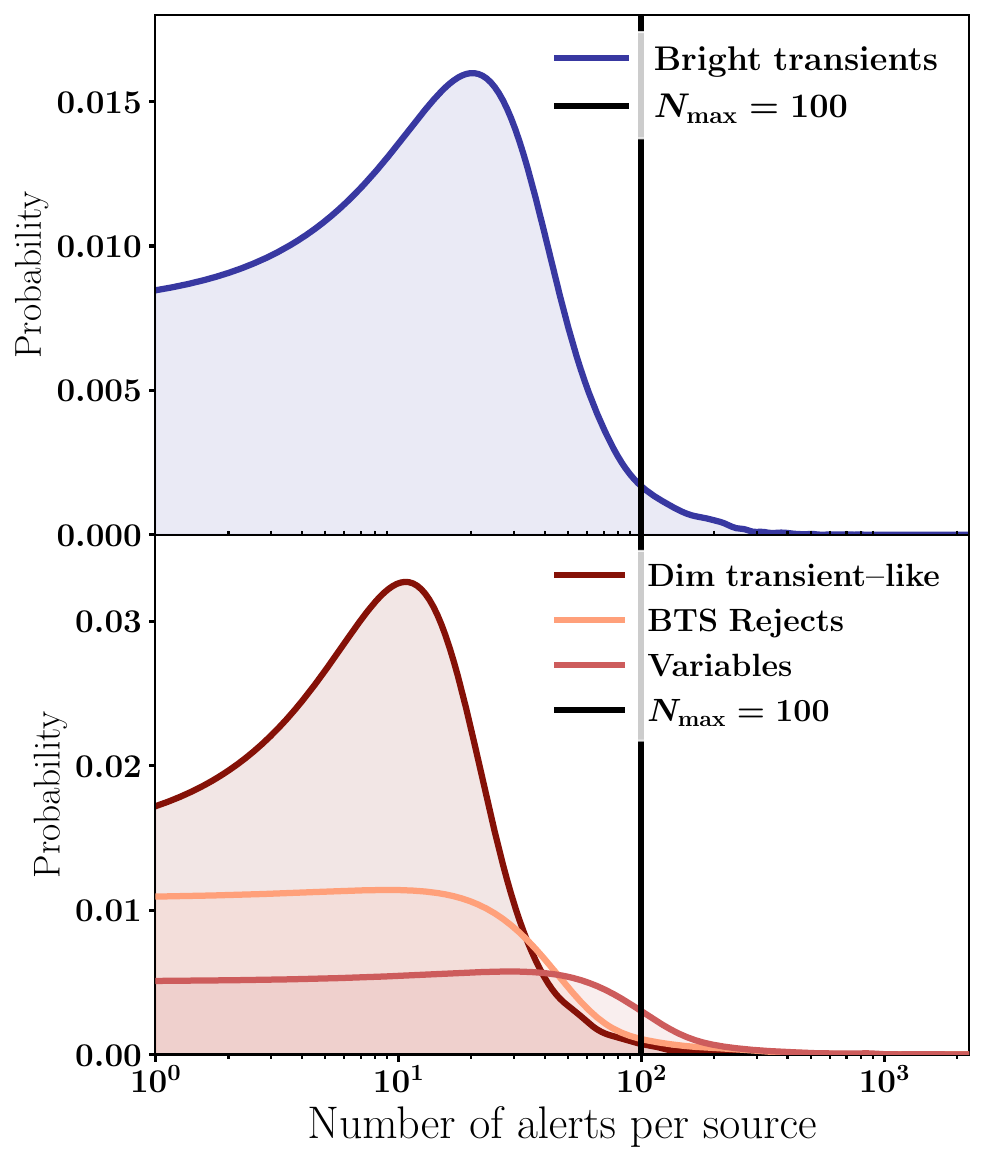}
    \caption{Gaussian kernel density estimations showing the number of alerts per source for each query comprising our cleaned training set. \textit{Top}: positive class examples; \textit{Bottom}: negative class examples. Sources with many alerts are thinned down to $N_\mathrm{max}=100$ alerts per source. This prevents long-lived sources like AGN (in Variables) and bright SNe (in Bright transients) from being over-represented in training.}
    \label{fig:Nmax}
    \end{center}
\end{figure}

Figure~\ref{fig:Nmax} illustrates that, as expected, the typical number of alerts per source is variable over the four different queries. Some AGN in \textit{vars} have $>10^{3}$ alerts, while some SNe have as few as two alerts. We correct this imbalance by defining a hyperparameter $N_{\mathrm{max}}$: the maximum number of alerts per source allowed in the training set. Sources with more than $N_{\mathrm{max}}$ alerts will have some alerts removed and those with $N_{\mathrm{max}}$ or fewer alerts will remain untouched. 

\begin{figure*}[ht]
    \begin{center}
    \includegraphics[width=2\columnwidth]{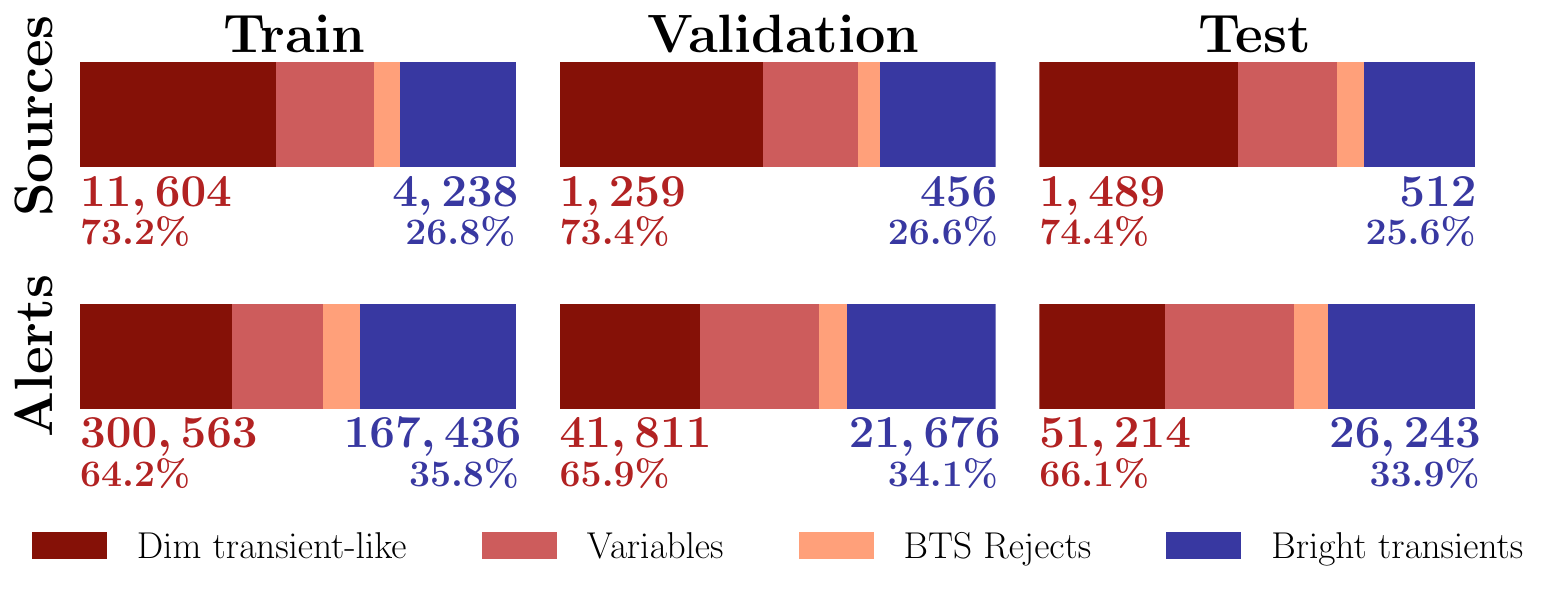}
    \caption{Bar charts showing the distribution of sources and alerts from 4 queries into train/validation/test (81\%/9\%/10\%) splits. Certain sources have an upper limit set on their number of alerts, excess alerts are removed. Our training set is imbalanced favoring not-BTS sources $\sim$3:1 and favoring the not-BTS alerts $\sim$2:1. }
    \label{fig:trainvaltest}
    \end{center}
\end{figure*}

We apply $N_{\mathrm{max}}$ alert thinning to sources based on which query they originate from and which split they are present in. For sources in the \textit{trues}, \textit{dims}, and \textit{rejects} queries (many of which are galactic or extragalactic transients), we want \BTSbot~to learn and identify their properties at all phases of their evolution, so we take a simple uniform random selection of $N_{\mathrm{max}}$ alerts to preserve and discard all others. The remaining alerts are roughly uniformly distributed over the rise, peak, and fade of these transients\footnote{Observational coverage of transients is typically greater near peak, so this simple random selection will slightly favor removing alerts near peak. We do not expect significantly different results if we attempted to correct for this.}. For sources in the \textit{vars} query, we keep only the most recent $N_{\mathrm{max}}$ alerts. Metadata of long-lived sources look systematically different in the present than they did years ago. For example, AGN were typically first detected near the beginning of ZTF and have been repeatedly detected since. New incoming alerts from these AGN, those that \BTSbot~will encounter in production, will typically have very large \texttt{age} and \texttt{ndethist} (approximately, number of previous detections), but those nearer to when they were first detected will have much smaller \texttt{age} and \texttt{ndethist}. By selecting the latest $N_{\mathrm{max}}$ alerts, we capture those which are most representative of today's BTS candidates, and we avoid a data shift that would dramatically worsen \BTSbot's performance in production. Early versions of \BTSbot~used random $N_{\mathrm{max}}$ thinning on \textit{vars} sources and yielded markedly worse performance. We find the optimal value to be $N_{\mathrm{max}}=100$ (see Sec.~\ref{sec:train_and_opt} for optimization details).

The CVs in \textit{vars} are transients like the sources in \textit{trues} and \textit{dims}, and we want to capture alerts from all stages of their evolution, rather than just late-time. The selection of the latest $N_{\mathrm{max}}=100$ alerts from \textit{vars} sources does not significantly impact our coverage of CV evolution because most CVs are very rapidly evolving and have fewer than $N_{\mathrm{max}}=100$ alerts, so they typically receive no thinning at all. 

Sources from all queries in the train split get $N_{\mathrm{max}}$ alert thinning to avoid the over-representation problem. In the validation and test splits, however, only sources from the \textit{vars} query get alert thinning. The other sources are left un-thinned because (i) over-representation is not an issue, and (ii) we need all alerts from these sources to most accurately measure completeness and purity of the \BTSbot~predictions. We keep alert thinning in place for the \textit{vars} sources in the validation and test splits because the very old alerts from these long-lived AGN are unlike what appears in present day scanning.

Figure~\ref{fig:trainvaltest} shows the number of sources and alerts in each split and the query which they originated from after all cuts and thinning are applied. Our two classes are moderately imbalanced: $\sim$35\% BTS alerts to $\sim$65\% not-BTS alerts. We discuss techniques for mitigating the effects of class imbalance in Sec.~\ref{sec:architecture&training}. In total, our production training set comprises $608\mathrm{,}943$ total alerts from $19\mathrm{,}558$ total sources totalling $>$60\,GB. Despite \BTSbot's relatively narrow domain, this is significantly more alerts than other models with similar architectures, such as the ALeRCE stamp classifier \citep[$\sim$52,000;][]{Carrasco-Davis+2021} and the ACAI models \citep[$\sim$200,000;][]{Duev+2021}.

%% /------------------------------------/
%% /-------- MODEL ARCHITECTURE --------/ 
%% /------------------------------------/
\section{\BTSbot~Scope, Architecture, and Training} 
\label{sec:architecture&training}

% Automated alert-based identification typically expedites this process with respect to the baseline defined by the BTS human scanners. In this way, \BTSbot~helps to maintain BTS's near-perfect completeness by obtaining a spectrum before some transient fades beyond detection.

\begin{figure*}[ht!]    \centerline{\includegraphics[width=1.9\columnwidth]{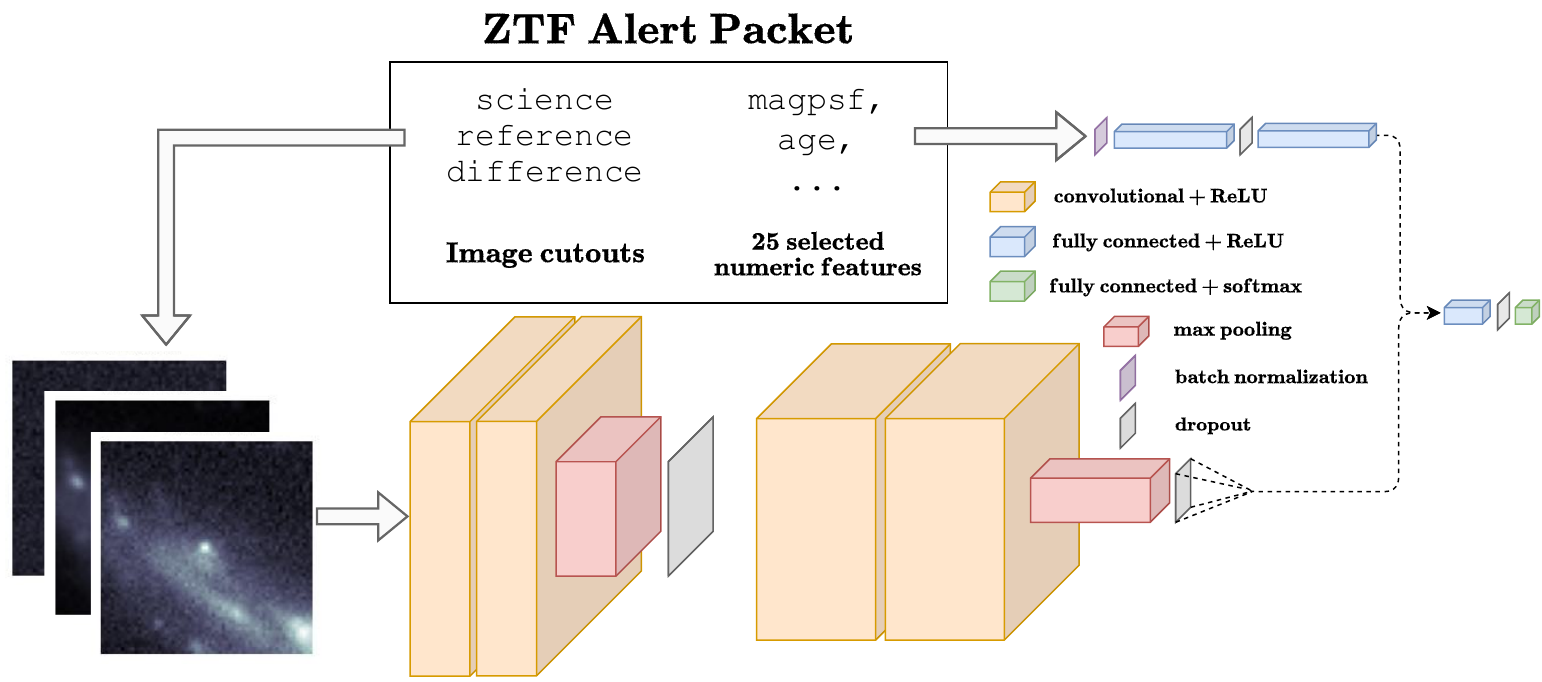}}
    \caption{Diagram of our multi-modal convolutional neural network, \BTSbot, performing bright extragalactic transient / not-bright extragalactic transient binary classification on ZTF alert packets. Image input is processed through the convolutional branch and then flattened to a 1D vector and concatenated with the output of dense layers in the metadata branch. After another dense layer, a single-neuron layer produces the final prediction: a unit-interval bright transient score.}
    \label{fig:model_diagram}
\end{figure*}

Our MM-CNN, \BTSbot, automates source identification for BTS by assigning each ZTF alert packet a bright transient score. Figure~\ref{fig:model_diagram} shows how input is fed into \BTSbot~and how information is combined to produce the output score; \BTSbot~contains three main components. (1) The convolutional branch processes the science, reference, and difference cutouts as a three-channel image through a VGG-like architecture \citep{Simonyan+2014}. (2) The metadata branch processes the 25 extracted features (see Table~\ref{tab:metadata_features}) through a batch normalization layer and two dense layers. (3) The combined section concatenates the output of the two branches and passes it through two more dense layers, the second of which produces the prediction using a softmax activation function. The output is a unit-interval score where higher scores represent increased confidence that the source in the input alert packet is, or will become, a bright extragalactic transient. The parameters for each layer are shown in Table~\ref{tab:model_arch}. Other than the final layer, activation functions are all the rectified linear unit \citep[ReLU;][]{Nair&Hinton2010}, and all convolutional layers have symmetric unit stride and same padding.

\begin{deluxetable}{cc|c}
\tablecaption{\BTSbot~layer configurations \label{tab:model_arch}}
\tablehead{
    \colhead{Layer type} & \colhead{Layer parameters} & \colhead{\makecell{Hyperparameter \\ search range}}
}
\startdata
    \multicolumn{2}{c}{Convolutional branch} & \\
    \hline
    2D Conv. & 32 filters, $5\times5$~kernel & 8\,--\,128 filters\tablenotemark{a} \\
    2D Conv. & 32 filters, $5\times5$~kernel & [3, 5, 7] kernel size\tablenotemark{a} \\
    Max pool & $2\times2$~kernel & - \\
    Dropout & 0.50 & 0.1\,--\,0.8 \\
    2D Conv. & 64 filters, $5\times5$~kernel & 8\,--\,128 filters\tablenotemark{a} \\
    2D Conv. & 64 filters, $5\times5$~kernel & [3, 5, 7] kernel size\tablenotemark{a} \\
    Max pool & $4\times4$~kernel & - \\
    Dropout & 0.55 & 0.1\,--\,0.8  \\
    % \hline
    \hline
    \multicolumn{2}{c}{Metadata branch} & \\
    \hline
    Batch norm. & - & - \\
    Dense & 128 units & 32\,--\,256 units \\
    Dropout & 0.25 & 0.1\,--\,0.8 \\
    Dense & 128 units & 32\,--\,256 units \\
    % \hline
    \hline
    \multicolumn{2}{c}{Combined section} & \\
    \hline
    Dense & 8 units & 8\,--\,128 units  \\
    Dropout & 0.20 & 0.1\,--\,0.8 \\
    Dense & 1 unit & - \\
    \hline
\enddata
\tablenotetext{a}{All 2D Convolutional (Conv.) layers have the same search range for filter counts and kernel size.}
\tablecomments{Images and metadata are passed through their respective branches, and the output of either is concatenated and sent to the combined branch. Dropout and batch normalization (batch norm.) layers are included to regularize.}
\end{deluxetable}

The choice of a MM-CNN is motivated by the fact that the images and the extracted features provide complementary information for performing our task. For example, the extracted feature \texttt{distpsnr1} represents the angular distance to the PS1 cataloged source nearest to this ZTF source, and \texttt{sgscore1} represents the star-galaxy score \citep{Tachibana_Miller_2019} of this PS1 source. While new transients are not present in PS1, the host galaxies of supernovae (SNe) often are. Thus, alerts from SNe tend to have moderate \texttt{distpsnr1} and small \texttt{sgscore1} values, indicating a galaxy projected nearby to the source. Most AGN and some CVs that pass the BTS alert filter are cataloged in PS1 and thus have \texttt{distpsnr1} very near to zero. The images also provide important information following a similar heuristic. Bright SNe tend to be associated with prominent (bright with large angular size) off-center extended sources, their host galaxies; faint SNe tend to have less prominent host galaxies because they tend to be farther away; AGN will appear as exactly centered extended sources; CVs will often appear surrounded by many bright point sources because they tend to occur near the galactic plane around many other stars. \cite{Carrasco-Davis+2021} give more detailed descriptions of how AGN, SNe, and VarStars can be distinguished using ZTF image cutouts and metadata. A MM-CNN is able to pool information from all input types and consider them together when making a prediction. We also experiment with uni-modal alternatives to \BTSbot~in Appendix~\ref{app:unimodal}. 

Given the scope and architecture of \BTSbot~we encounter a number of challenges. First, we are requiring \BTSbot~to learn multiple complex separations. \BTSbot~must learn to separate SNe from other sources without using redshift information because it is not always known \textit{a priori}. It must also learn to identify bright SNe with limited time-series information irrespective of the SN's current phase. Early in its rise or late in its fade, a bright SN can appear very similar to a near-peak dim SN. Additionally, \BTSbot~does not have explicit information stating the BTS threshold is 18.5 mag. Rather, it must learn the position of this threshold drawn through the continuous distribution of SNe peak magnitudes.
Lastly, \BTSbot~does not have direct access to full light curve information, i.e., information on every previous individual detection. Instead, it has access to the basic custom metadata features we compute from the full light curve. We omit full light curve information, in part, because there is no established method in the literature for representing partial light curves of the wide variety our model encounters in a way that is fit for input into a neural network. There has been a great deal of work to accomplish this for SNe alone \citep[e.g.,][]{Villar+2020}, but these methods are not applicable to all the types of sources that \BTSbot~encounters. This choice supports the possibility of tuning \BTSbot~to identify extragalactic transients very rapidly, potentially on their first detection, which Section~\ref{sec:2023ixf} explores.

\subsection{Training and hyperparameter optimization}
\label{sec:train_and_opt}

\BTSbot~is implemented with \texttt{TensorFlow} \citep{tensorflow} and the Keras API \citep{keras}. We adopt the Adam optimizer \citep{Kingma+2014} and the binary cross-entropy loss function. In addition to thinning a source's alerts by $N_\mathrm{max}$ (see Sec.~\ref{sec:train_data}), we make use of a number of training techniques to mitigate overfitting. We employ data augmentation, which executes random rotations of 0$^\circ$, 90$^\circ$, 180$^\circ$, and 270$^\circ$ and random horizontal and vertical flipping on the image cutouts. These also help ensure that \BTSbot~is invariant to these transformations. We weight contributions to the loss function by the inverse of the relative size of the input alert's class (i.e. misclassifications of BTS alerts contribute more loss than that of not-BTS alerts). This size is computed by the count of alerts per class in the train split, but we also experiment with computing these weights based on the number of sources per class. The learning rate $\alpha$ decreases after 20 epochs without improved validation loss and training terminates after 75 epochs without improved validation loss. 

\begin{deluxetable}{cc|c}
\tablecaption{\BTSbot~hyperparameters \label{tab:hyperparams}}
\tablehead{
    \colhead{Parameter name} & \colhead{Optimized value} & \colhead{\makecell{Hyperparameter \\ search range}}
}
\startdata
    batch size & 64                           & 8\,--\,64 \\
    Adam $\beta_1$ & 0.99                     & 0.81\,--\,0.999 \\
    Adam $\beta_2$ & 0.99                     & 0.9\,--\,0.9999 \\
    learning rate ($\alpha$) & $10^{-4}$      & $10^{-2}$\,--\,$5\times10^{-6}$ \\
    $\alpha$ decrease factor & $0.4$          & 0.25\,--\,0.75 \\
    $\alpha_\mathrm{min}$ & $5\times10^{-10}$ & $10^{-10}$\,--\,$10^{-5}$ \\
    $N_\mathrm{max}$ & 100                    & 1\,--\,$\infty$ \\
\enddata
\end{deluxetable}

\begin{deluxetable}{l|c}
\tablecaption{\BTSbot~metadata features \label{tab:metadata_features}}
\tablehead{
    \colhead{Feature name} & \colhead{Definition [unit]}
}
\startdata
    \multicolumn{2}{c}{Alert packet metadata} \\
    \hline
    \texttt{sgscore\{1,2\}}  & Star/Galaxy score of nearest two PS1 sources \\
    \texttt{distpsnr\{1,2\}} & Distance to nearest two PS1 sources [arcsec] \\
    \texttt{fwhm}            & Full Width Half Max [pixels] \\
    \texttt{magpsf}          & magnitude of PSF-fit photometry [mag] \\
    \texttt{sigmapsf}        & 1-$\sigma$ uncertainty in \texttt{magpsf} [mag] \\
    \texttt{chipsf}          & Reduced $\chi^2$ of PSF-fit \\
    \texttt{ra}              & Right ascension of source [deg] \\
    \texttt{dec}             & Declination of source [deg] \\
    \texttt{diffmaglim}      & 5-$\sigma$ magnitude detection threshold [mag] \\
    \texttt{ndethist}        & Number of previous detections of source \\
    \texttt{nmtchps}         & $\#$ of PS1 cross-matches within 30 arcsec \\
    \texttt{drb}             & Deep learning-based real/bogus score \\
    \texttt{ncovhist}        & $\#$ of times source on a field and read channel \\
    \texttt{chinr}           & $\chi$ parameter of nearest source in reference \\
    \texttt{sharpnr}         & sharp parameter of nearest source in reference \\
    \texttt{scorr}           & Peak-pixel S/N in detection image \\
    \texttt{sky}             & Local sky background estimate [DN] \\
    \hline
    \multicolumn{2}{c}{Custom metadata} \\
    \hline
    \texttt{days\_since\_peak} & Time since brightest alert [days] \\
    \texttt{days\_to\_peak} & Time from first to brightest alert [days] \\
    \texttt{age} & \texttt{days\_since\_peak} + \texttt{days\_to\_peak} \\
    \texttt{peakmag\_so\_far} & Source's minimum \texttt{magpsf} thusfar [mag] \\
    \texttt{maxmag\_so\_far} & Source's maximum \texttt{magpsf} thusfar [mag] \\
    \texttt{nnondet}\tablenotemark{a} & \texttt{ncovhist} - \texttt{ndethist} 
    % \hline
    % \multicolumn{2}{c}{Tested but excluded} \\
    % \hline
    % \texttt{sgscore3} & - \\
    % \texttt{distpsnr3} & - \\
    % \texttt{fid}
    % \texttt{maggaia} & - \\
    % \texttt{neargaia} & - \\
    % \texttt{magdiff} & - \\
    % \texttt{magap} & - \\
    % \texttt{sigmaap} & - \\
    % \texttt{magapbig} & - \\
    % \texttt{sigmaapbig} & - \\
    % \texttt{magnr} & - \\
    % \texttt{ssnrms} & - \\
    % \texttt{dsnrms} & - \\
    % \texttt{seeratio} & - \\
    % \texttt{nneg} & - \\
    % \texttt{magzpsci} & - \\
    % \texttt{jdstarthist} & - \\
    % \texttt{classtar} & - \\
\enddata
\tablenotetext{a}{Adopted from \cite{Carrasco-Davis+2021}.}
\tablecomments{The 25 metadata features passed into \BTSbot's metadata branch. Full definitions of alert packet features can be found at \url{https://zwickytransientfacility.github.io/ztf-avro-alert/schema.html}.}
\end{deluxetable}

We utilize the Weights and Biases platform \citep{wandb} to perform multiple extensive Bayesian hyperparameter sweeps. A Bayesian hyperparameter sweep guides searching through a large grid of hyperparameters intelligently by leveraging correlations between previous hyperparameter inputs and the corresponding trained model's performance in a specified metric. This is expected to be more efficient than traditional grid search sweeps because, for example, the Bayesian sweep can learn to avoid certain regions of the hyperparameter space that frequently yield poor performing models. We find the optimal configuration of hyperparameters for \BTSbot~layers (see Table~\ref{tab:model_arch}), the selection of metadata features (see Table~\ref{tab:metadata_features}), and other values (see Table~\ref{tab:hyperparams}). Tables~\ref{tab:model_arch}~and~\ref{tab:hyperparams} also include the range of values searched over for each hyperparameter in our sweeps. The search over which alert packet metadata features to provide to \BTSbot~included many other features which we found to not improve performance: \texttt{sgscore3}, \texttt{distpsnr3}, \texttt{fid}, \texttt{maggaia}, \texttt{neargaia}, \texttt{magdiff}, \texttt{magap}, \texttt{sigmaap}, \texttt{magapbig}, \texttt{sigmaapbig}, \texttt{magnr}, \texttt{ssnrms}, \texttt{dsnrms}, \texttt{seeratio}, \texttt{nneg}, \texttt{magzpsci}, \texttt{jdstarthist}, and \texttt{classtar}.

%% /-----------------------------------/
%% /-------- MODEL PERFORMANCE --------/
%% /-----------------------------------/
\section{\BTSbot~performance and comparison to human scanners} 
\label{sec:4}

\subsection{Performance on the test split}
\label{sec:test_performance}

\begin{figure}
    \begin{center}
    \includegraphics[width=0.9\columnwidth]{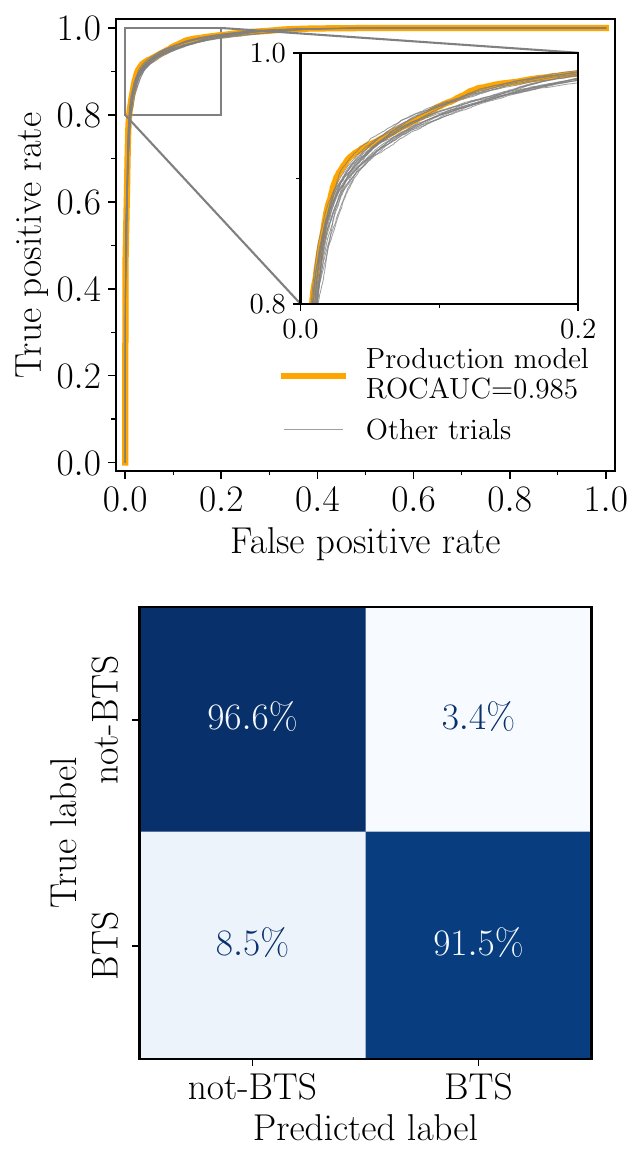}
    \caption{ROC curve (\textit{top}) and confusion matrix (\textit{bottom}) of the \BTSbot~production model and 19 other trials with different initializations. \textit{Top}: The production \BTSbot~model (orange curve) and 19 other trials (gray curves) yield very similar ROC curves and ROCAUCs, indicating that \BTSbot's excellent performance is robust to different initializations. \textit{Bottom}: The production model's confusion matrix makes clear that \BTSbot~has greater TN rate (upper-left quadrant) than TP rate (lower-right quadrant). We find that models which trade-off TP rate in favor of TN rate tend to improve in other key performance metrics.}
    \label{fig:roc_cm}
    \end{center}
\end{figure}

To characterize \BTSbot's generalizable performance, we present performance metrics on test split data. We train 20 trials of \BTSbot~with identical optimal hyperparameters and different random initializations of the model's learnable parameters. The trial which yields the best test split performance is chosen as the production model. Performance metrics are reported for the chosen production model and the median of the metric for the 20 trials with uncertainties representing the metric's 1-$\sigma$ bounds.

The final \BTSbot~models yield test accuracy of $94.1\%\pm0.28$ and \testacc~for the production model. The upper panel of Figure~\ref{fig:roc_cm} shows \BTSbot's receiver operating characteristic (ROC) curves. ROC curves visualize the balance of true and false positives at various classification thresholds, and the area under the ROC curve (ROCAUC) is frequently used as a summary statistic for the performance of a classifier. A perfect classifier's ROC curve has ROCAUC equal to unity while ROCAUC=0.5 corresponds to random guessing (on a balanced dataset). The final \BTSbot~models yield ROCAUC=$0.984\pm0.001$ and ROCAUC=\testrocauc~for the production model. The results of the production model and the 19 other trials are extremely similar, indicating stability in \BTSbot's performance over different initializations. The lower panel of Figure~\ref{fig:roc_cm} shows the production model's confusion matrix. A binary classifier's confusion matrix visualizes the frequencies of the four classification outcomes: true negative ($\mathrm{TN}$), false positive ($\mathrm{FP}$), false negative ($\mathrm{FN}$), and true positive ($\mathrm{TP}$). The production model shows higher accuracy on not-BTS alerts (TN rate) than on BTS alerts (TP rate). We could attempt to train \BTSbot~to better balance the TP and TN rates, for example by adjusting the class-weights, but we find that models which favor TN rate perform better overall.

While we fare excellently in traditional ML performance metrics, they are not necessarily representative of \BTSbot's real-world performance. Because \BTSbot~produces scores on alert packets (alert-based classification) but sources must be chosen for follow-up (source-based classification), we must define a mapping from a sequence of alert predictions to a source prediction. We call these mappings ``policies," which we define to be analogous to the criteria BTS scanners use when deciding whether or not to save a source and request a spectrum of it. By simulating our policies on BTS candidates, we can compute performance metrics which realistically represent how \BTSbot~performs as a scanner. Further, we can compare the resulting figures against human scanners to contextualize \BTSbot's performance. 

We define two policies which closely emulate their human scanning analogs: $\texttt{bts\_p1}$ and $\texttt{bts\_p2}$. The policy $\texttt{bts\_p1}$ requires that a source have at least two alerts with high ($\geq0.5$) bright transient score and $\texttt{magpsf}\leq19~\mathrm{mag}$ before being saved and having an SEDM trigger sent at priority 1. The policy $\texttt{bts\_p2}$ requires that a source meet $\texttt{bts\_p1}$ as well as having at least one alert with $\texttt{magpsf}\leq18.5~\mathrm{mag}$ before a trigger being sent with priority 2. Priority is a parameter of the requests sent to SEDM, where larger values indicate the request is more urgent. Higher priority requests are typically fulfilled before lower priority requests, although other factors, e.g. observability, are also taken into account. Most follow-up requests sent by BTS scanners are with priority 1 or 2, and priorities 2 and greater are typically reserved for sources which have already attained $m_\mathrm{peak}\leq18.5~\mathrm{mag}$. 

\begin{figure*}[ht]
    \begin{center}
    \includegraphics[width=1.8\columnwidth]{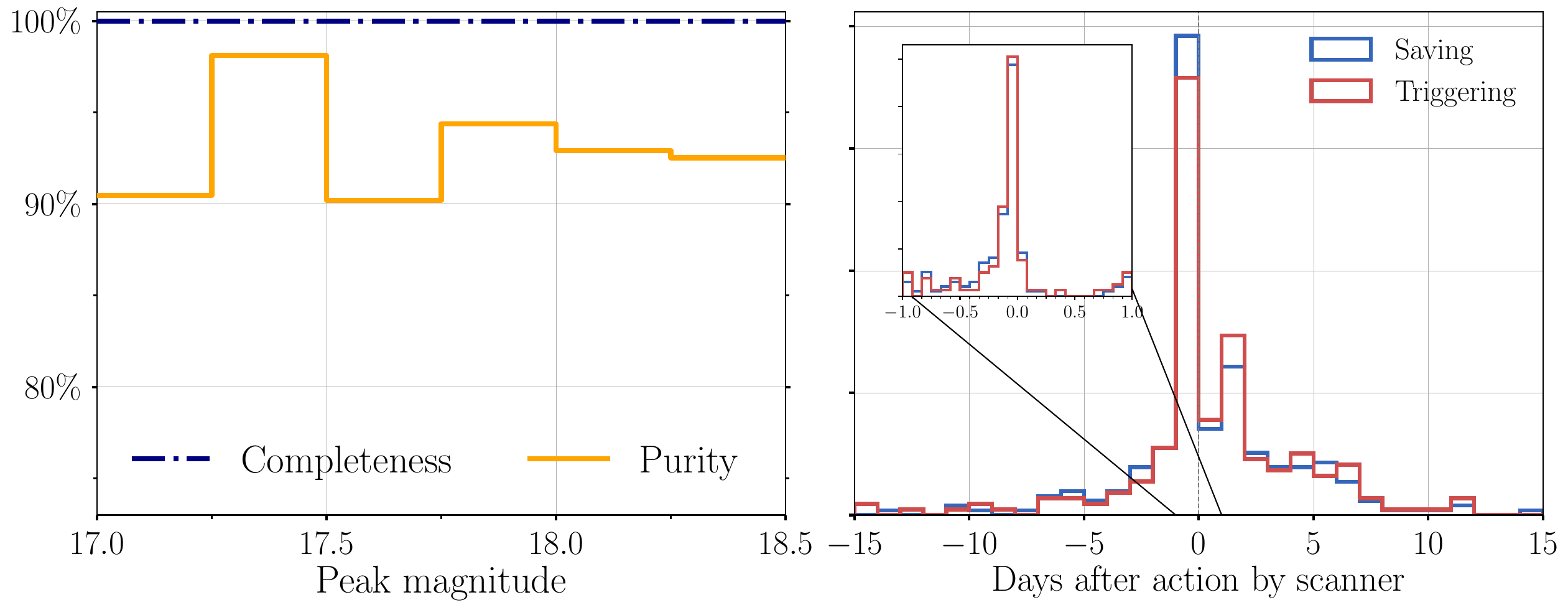}
    \caption{Completeness and purity of \BTSbot~actions (\textit{left}) and speed comparison with human scanners (\textit{right}) for sources in our test split. \textit{Left}: The completeness curve (dash-dot navy) is 100\% in all bins. \BTSbot's perfect completeness is conducive to BTS's science efforts which require a highly-complete, unbiased sample. The small variations in the purity curve (solid orange) are due to small number statistics. The overall purity of $\texttt{bts\_p1}$ and $\texttt{bts\_p2}$ are \ponetestpurity~and \ptwotestpurity~respectively. $\texttt{bts\_p1}$ has extra contamination from SNe with $m_\mathrm{peak}$ slightly greater than 18.5 mag, sources which are acceptable targets for spectroscopic observation. \textit{Right}: Histograms comparing \BTSbot's speed in saving (blue) and triggering (red) with that of scanners. Both distributions peak very near to zero indicating that \BTSbot~acts as quickly as human scanners on new bright transients.
    }
    \label{fig:test_performance}
    \end{center}
\end{figure*}

We quantify the performance of these policies with completeness, purity, $\Delta t_\mathrm{save}$, and $\Delta t_\mathrm{trigger}$. Completeness (or ``recall") is the fraction of bright transients which are correctly classified by the policy: $\mathrm{TP}/(\mathrm{TP} + \mathrm{FN})$. The completeness of $\texttt{bts\_p1}$ and $\texttt{bts\_p2}$ are identical, so we report a single value for both policies. Purity (or ``precision") is the fraction of predicted bright transients that are actually bright transients: $\mathrm{TP}/(\mathrm{TP} + \mathrm{FP})$. $\Delta t_\mathrm{save/trigger}$, is the difference between the Julian Date ($\mathrm{JD}$) at which \BTSbot~saved and triggered on a source with the $\mathrm{JD}$ that scanners did the same.  The $\mathrm{JD}$s for scanners are queried from Fritz for all sources in the \textit{trues} set (see Sec.~\ref{sec:train_data}). Sources that were saved before January 1st 2021 are removed from this analysis because many of them were scanned using the GROWTH Marshal \citep{Kasliwal+2019}, so the save and trigger $\mathrm{JD}$s available on Fritz are unreliable. For saving, we use the JD at which a scanner added a source to the BTS catalog on Fritz. For triggering, we only consider sources which had an SEDM integral field unit (IFU) request sent before their first spectrum was uploaded to Fritz. The JD used for comparison is the JD at which the first IFU follow-up request was created. This restriction is applied because the time at which scanners decided to trigger is not available for the other facilities BTS uses for classification. \BTSbot's $\mathrm{JD}$ for saving and triggering on a source is the JD associated with the alert that first made the source satisfy $\texttt{bts\_p1}$. The \BTSbot~JDs correspond to either policy, but we select \texttt{bts\_p1} as it makes for a more direct comparison to scanners.

% Completeness and purity are chosen because they are crucial to BTS's science. The primary mission of BTS is to spectroscopically classify \textit{all} transients passing their cuts. Imperfect completeness results in a potentially biased BTS sample and one that yields less precise constraints on transient populations. Imperfect purity in saving makes monitoring of the survey less efficient, and imperfect purity in triggering results in wasted spectroscopic time. \nr{Add something like "These spectra, usually of CVs, are shared with the ZTF members who do make use of them."?} 

When computing completeness and purity for these policies, we make two additional minor cuts. First, 70 sources in the test split also appear in \textit{junk} (see Sec.~\ref{sec:train_data}). This catalog is a list of sources that are unambiguously not bright transients which frequently pass the BTS alert filter. When scanning, they are typically hidden and not considered for saving or triggering; they are excluded from completeness and purity calculations. We also identify 59 sources which have only a single alert remaining after the cleaning cuts (see Sec.~\ref{sec:train_data}). These sources will never pass either policy, so they are also excluded from completeness and purity calculations. 

We use estimates of the scanners' completeness, purity, and speed as benchmarks against which to judge \BTSbot's performance. We compute a lower-limit on the scanners' saving completeness of bright transients from values presented in Table~1 and Section~3 of \cite{Perley+2020}, which give 99.6\% completeness.
% The completeness values we present for the BTS sample or \BTSbot~are always defined as a fraction of bright transients passing the BTS alert filter or bright transients in the input data set. There are a very small number of instances where bright transients do not pass the BTS alert filter, e.g. due to a misclassification in \texttt{drb} or because they are projected very near to a foreground star, and thus are not classified by BTS. 
It is not straightforward to compute the BTS scanners' purity for saving or triggering on bright transients. Scanners will often intentionally act on transients which they know will not attain $m_\mathrm{peak}\,\leq\,18.5\,\mathrm{mag}$,\footnote{These sources, while not ``bright" transients, are still of interest to BTS and the individuals within BTS.} so straightforward estimates of the scanners' bright transient purity would underestimate their ability to select only bright transients. Instead, we estimate their saving and triggering purity for selecting any extragalactic transient and rejecting other sources. We limit this analysis to saves and triggers performed between $2460175.5 < \mathrm{JD} < 2460216.5$ (see Sec.~\ref{sec:real_time_performance} for explanation of JD bounds). During this time, 266 sources were saved by scanners to the primary internal BTS catalog, only 4 of which were not extragalactic transients: $98.5\%$ scanner saving purity. Similarly, scanners sent SEDM IFU for 327 unique sources of which 11 were non-extragalactic transients: $96.7\%$ scanner triggering purity. Further, scanners saving and triggering on transients with $m_\mathrm{peak}>18.5~\mathrm{mag}$ complicates $\Delta t_\mathrm{save/trigger}$ estimates. Scanners will save and trigger on transients slightly faster than otherwise because they need not wait for a transient to unambiguously demonstrate that it will soon have $m_\mathrm{peak}\leq18.5~\mathrm{mag}$. In addition, BTS scanners are occasionally aided in identifying bright transients by TNS reports from scanners working with other surveys or other ZTF alert brokers. For these reasons, $\Delta t_\mathrm{save/trigger}$ comparisons will not be perfectly direct, but we still adopt them as a basic benchmark for \BTSbot's speed. 

\begin{figure*}[ht]
    \begin{center}
    \includegraphics[width=2\columnwidth]{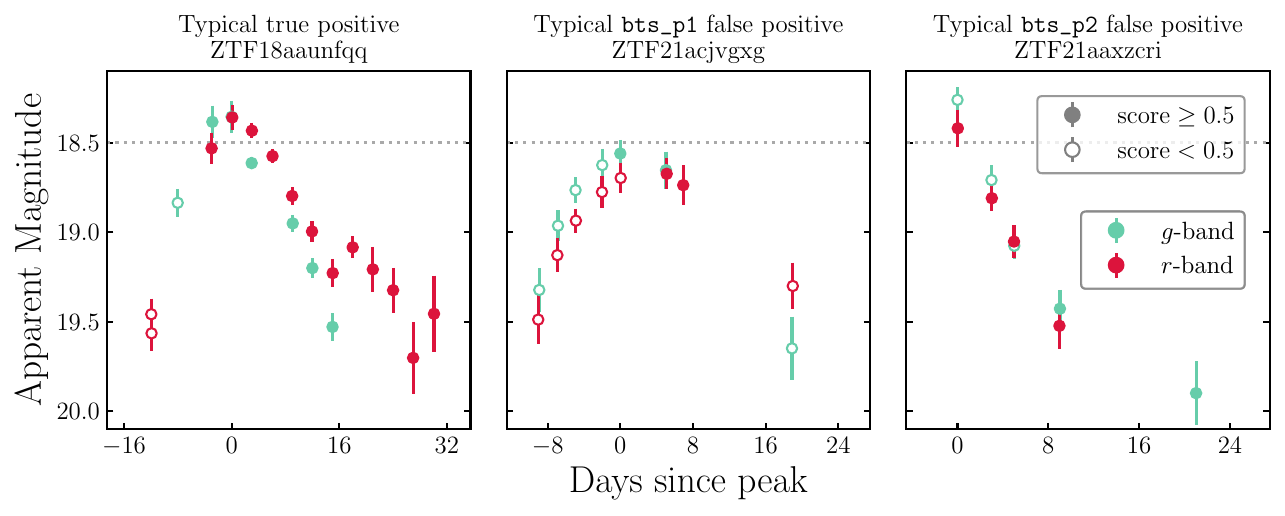}
    \caption{Light curves of three sources depicting typical evolution of \BTSbot~scores. Teal ($g$-band) and red ($r$-band) points indicate detections, and filled and open circles represent alerts which received score $\geq0.5$ and $<0.5$ respectively.  
    \textit{Left:} TPs may have low scoring alerts while still dim, but scores increase once they near the 18.5 mag threshold (dotted gray line). After fading well below the peak magnitude, scores remain high, in part, due to information provided by custom metadata features.
    \textit{Center:} Almost all \texttt{bts\_p1} FPs are dim ($m_\mathrm{peak}>18.5\,\mathrm{mag}$) transients whose alerts receive high scores when near the BTS threshold. 
    \textit{Right:} Many \texttt{bts\_p2} FPs are CVs, which could be better rejected by increasing the score threshold to $0.8$. 
    % Multiple of \BTSbot's~few false negatives are sources with very poor light curve coverage, giving \BTSbot~fewer opportunities to correctly select these bright transients.
    }
    \label{fig:LCs}
    \end{center}
\end{figure*}

The left panel of Figure~\ref{fig:test_performance} shows \BTSbot's completeness and purity under our policies as a function of peak magnitude. The completeness curve is exactly 100\% in all peak magnitude bins, giving perfect overall completeness.
% This is despite the fact that the dimmer BTS sources tend to have fewer alerts than the brighter ones, i.e., fewer chances for \BTSbot~to correctly classify the source. 
% These transients are expected to be a source of confusion for \BTSbot~because it has no explicit information stating the BTS threshold is 18.5 mag. For instance, two SNe with $m_\mathrm{peak}=18.45$ and $m_\mathrm{peak}=18.55$ appear very similar to each other in many aspects, but they have opposite classifications according to \BTSbot. 
This compares well to the BTS scanners' saving completeness: 99.6\%. Very high completeness is essential for maintaining the quality of the BTS sample and ensuring that \BTSbot~does not inject any significant selection biases into the BTS sample. The purity curve is $>90\%$ in all peak magnitude bins. Most bins have $<10$ FPs, so the variations between bins is likely due to small number statistics. Overall purity for $\texttt{bts\_p2}$ is \ptwotestpurity, and overall purity for $\texttt{bts\_p1}$ is \ponetestpurity. The \texttt{bts\_p1} purity is not aligned with the purity curve because \texttt{bts\_p1} also considers a large number of sources with $m_\mathrm{peak}>18.5$.\footnote{Completeness and purity are not shown for $m_\mathrm{peak}>18.5$ because they are undefined and uniformly zero respectively.} Contamination from dim ($m_\mathrm{peak}>18.5$) transients is the reason why the \texttt{bts\_p1} purity is much less than the \texttt{bts\_p2} purity; they make up 53 of the 54 FPs unique to \texttt{bts\_p1}. These 53 transients have median $m_\mathrm{peak}$ of 18.57 mag and many received spectroscopic follow-up requests by BTS. They are still of interest to members of BTS, so they represent a reasonable use of follow-up resources. 
% They are still of interest and value to BTS, so these misclassifications are of little concern. 
% Overall \texttt{bts\_p1} purity is 93.5\% when considering selections of dim transients as TPs, which makes for a more direct comparison with scanner purity estimates. 
\BTSbot's purity falls slightly ($\sim2\%-6\%$) short of the scanners' (\btssavepurity~saving, \btstriggerpurity~triggering). 
% Completeness, where \BTSbot~performs perfectly on the test split, is paramount to BTS but purity is also key for best allocating BTS's valuable spectroscopic time. 
Likely at the cost of completeness or speed, alternative policies could be designed to more conservatively allocate spectroscopic resources, increasing \BTSbot's purity to compare favorably with scanners. \BTSbot's~behavior is dependent on \texttt{braai} and \texttt{sgscore} because they contribute to determining what sources pass the BTS alert filter: the only sources which \BTSbot~trains and triggers on.

% Identification of sources for BTS is time-sensitive. If delayed, the transient may fade beyond detection before having its spectrum collected, thus reducing the BTS sample's completeness. Additionally, observing the transient near peak is favored because absorption lines used for classification are usually strong then and doing so minimizes exposure time needed. 

The right panel of Figure~\ref{fig:test_performance} shows histograms comparing the time at which \BTSbot~and the human scanners saved or triggered on a source. Negative values indicate that \BTSbot~was faster, and positive values indicate that the scanners were faster. Both histograms peak sharply at 0 days, suggesting that scanners and \BTSbot~act on new transients at the same time. The median of $\Delta t_\mathrm{save}$ and $\Delta t_\mathrm{trigger}$ are $\testmeddtsave$~days and $\testmeddttrigger$~days respectively; \BTSbot~acts marginally quicker than scanners. Much of this performance is likely due to \BTSbot's decisions being made immediately as new alerts filter through the ZTF and BTS pipelines, although BTS does benefit from consistent real-time scanning thanks to members in European time-zones. 

The tails of this distribution include 6 (11) sources which are saved by \BTSbot~a week or more before (after) scanners did. Nuclear and hostless SNe make up most of the cases where \BTSbot~was faster. This suggests \BTSbot~is less hesitant to claim these challenging sources to be bright transients. Most of the cases where \BTSbot~is slower than scanners are slowly evolving SNe with a history of detections down to $\sim$20 mag. These suggest that scanners can better use the evolution prior to reaching 19 mag to identify transients, and \BTSbot~needs brighter detections to identify sources. Late identifications of these sorts are unlikely an issue because \BTSbot~still consistently identifies these sources before or near peak. Sources with poor light curve coverage, especially around 19 mag, cause large $\Delta t_\mathrm{save}$ and $\Delta t_\mathrm{trigger}$ and appear in either of these groups.
% Sources found quicker by BTSbot
Overall, \BTSbot~fares very well in time to save and trigger when compared to scanners.

% The blips of saves and triggers at +2 days is due to the 2-day cadence of the ZTF public survey; these are cases where \BTSbot~needed an additional night of observations to declare a source a bright transient.

% In some of these cases, scanners would likely wait for further detections if instructed to act exclusively on $m_\mathrm{peak}<18.5$ transients.

% The inset panel shows that a great number of sources have less than 6 hours of delay from being saved by~\BTSbot to being saved by human scanners.

% Scanners are able to consistently scan in real-time thanks to the BTS members located in European time-zones, where it is morning during the night at Palomar observatory. 

\subsubsection{Analysis of misclassifications in test split}
\label{sec:test_misclass}

Tracking and categorizing misclassifications is a key part of the development of ML models. Misclassifications are particularly important to understand in the case of \BTSbot~because mistakes could introduce biases into the BTS sample and waste valuable spectroscopic resources. 

On test split data, \texttt{bts\_p1} selects 92 FPs and 0 FNs. The majority of the FPs (53/92) are real dim transients. These are sources outside of \BTSbot's positive class but are still of interest to BTS, so they are acceptable FPs. Nearly all of the remaining FPs are CVs, AGN or QSOs and are shared with \texttt{bts\_p2}. The center panel of Figure~\ref{fig:LCs} presents an instance of a typical \texttt{bts\_p1} FP. Alerts are classified as not belonging to a bright transient until the source nears the 18.5 mag threshold, at which point a small number of high-scoring alerts cause the source to pass \texttt{bts\_p1}. The right panel shows a FP CV misclassified by both \texttt{bts\_p1} and \texttt{bts\_p2}. It shows high- and low-scoring alerts interspersed with each other, although some other CV FPs begin receiving exclusively low scores once the CV has faded well beyond peak. The completeness is perfect for \texttt{bts\_p1} and \texttt{bts\_p2}, so they have zero FNs. 
% Each of these 2 sources (ZTF18aayjflv, ZTF18abcryxz) are nuclear SNe with poor light curve coverage. ZTF18abcryxz additionally fails the quality cut defined in \cite{Perley+2020}, so it is not contained in the BTS statistical sample making it of less concern. The right panel of Figure~\ref{fig:LCs} presents the light curve of ZTF18abcryxz, which demonstrates the paucity of alerts for this source.
% that \BTSbot~had very few alerts to correctly identify this bright transient with.

Overall, the non-SN misclassifications are dominated by a relatively small number of CVs, AGN, and QSOs. Improved policies may be able to improve the rejection of these frequent FPs by, e.g., better leveraging existing information in external catalogs. The majority of the total misclassifications (dim SNe selected by \texttt{bts\_p1}) are not problematic for BTS and represent an appropriate allocation of spectroscopic time.

\subsection{Performance on very recent BTS candidates} 
\label{sec:real_time_performance}

Performance diagnostics computed on test split data tend to be robust and representative of real-world performance, but, in some cases, can have associated biases. Our test split includes many alerts that are years old and a subtle data shift (caused by, e.g., maintenance to the camera or the optics) may have occurred since then. To characterize \BTSbot's present-day performance we conduct an additional analysis using alerts from very recent BTS candidates. 

We perform this analysis in two parts: (i) we present alert-based performance metrics on alerts that recently passed the BTS alert filter and our cleaning cuts (see Sec.~\ref{sec:train_data}); (ii) we also present policy-based performance metrics on sources that recently passed $\texttt{bts\_p1}$ or recently peaked. The date boundaries for these analyses are determined by the final date our training data was queried (19 August 2023) and the date \BTSbot~went into production (29 September 2023): $2460175.5 < \mathrm{JD} < 2460216.5$. This cut on JD minimizes the bias in this analysis from transient alerts which \BTSbot~trained on and instances where \BTSbot's actions, which are visible to scanners, influenced scanners' decisions. Some very long-lived sources, like certain AGN, do have alerts in our training data and in this cleaned present-day sample. We do not remove these sources because they are encountered by \BTSbot~in production, and thus should be accounted for in this analysis.

\begin{figure}[ht]
    \begin{center}
    \includegraphics[width=0.9\columnwidth]{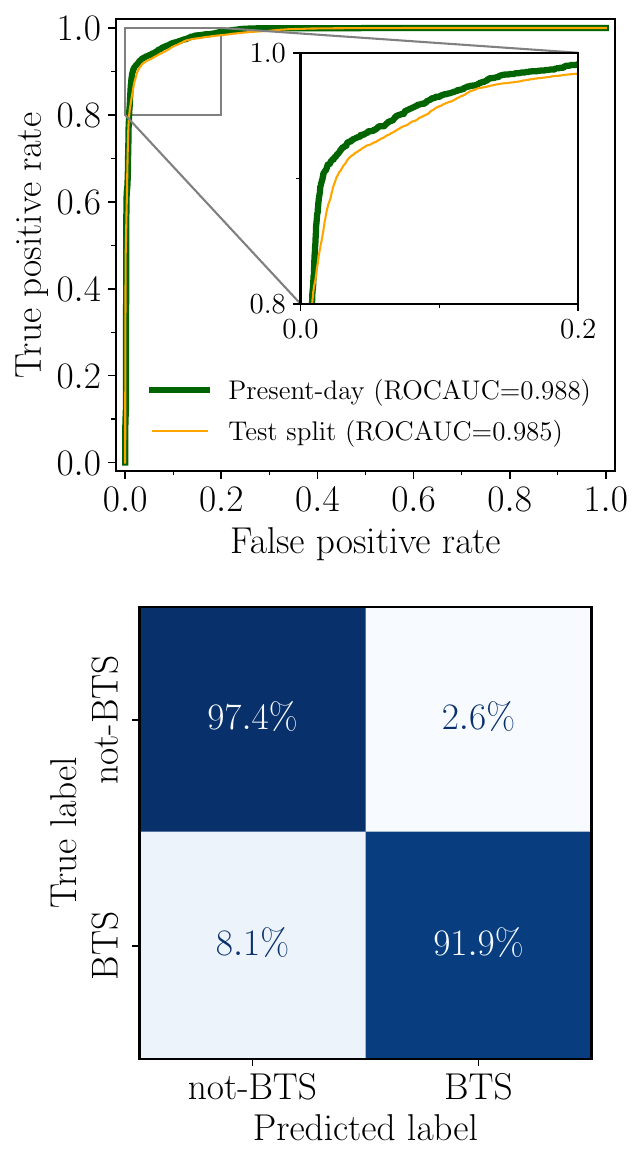}
    \caption{Same as Figure~\ref{fig:roc_cm} for a sample of alerts from very recent BTS candidates. The results of the ROC (present-day: thick green; test split: narrow orange), the ROCAUC, and the confusion matrix are all very similar to their test split analogs. All metrics are marginally improved for the present-day sample but not significantly so. These suggest that no data shift has occurred that significantly decreases \BTSbot's performance.}
    \label{fig:realtime_roc_cm}
    \end{center}
\end{figure}

We begin this analysis by applying the cleaning cuts described in Sec.~\ref{sec:train_data} on the public alerts that passed the BTS filter with $2460175.5 < \mathrm{JD} < 2460216.5$. After cuts, this sample totals 4,031 alerts from 251 bright transients and 15,159 alerts from 1,652 non-bright transients. Figure~\ref{fig:realtime_roc_cm} shows that the production \BTSbot~model yields \pdayacc~accuracy and ROCAUC=\pdayrocauc~on the present-day sample. TP rate and TN rate are 91.9\% and 97.4\%, respectively. The resulting performance is very similar to the metrics computed from test split data in Sec.~\ref{sec:test_performance} and shown in Figure~\ref{fig:test_performance}. Here, we observe marginally higher performance across all alert-based metrics than in the test split results. These variations are most likely due to the relatively small size of the present-day sample and are not indicative of a data shift affecting \BTSbot's performance.

\begin{figure*}[ht]
    \begin{center}
    \includegraphics[width=1.8\columnwidth]{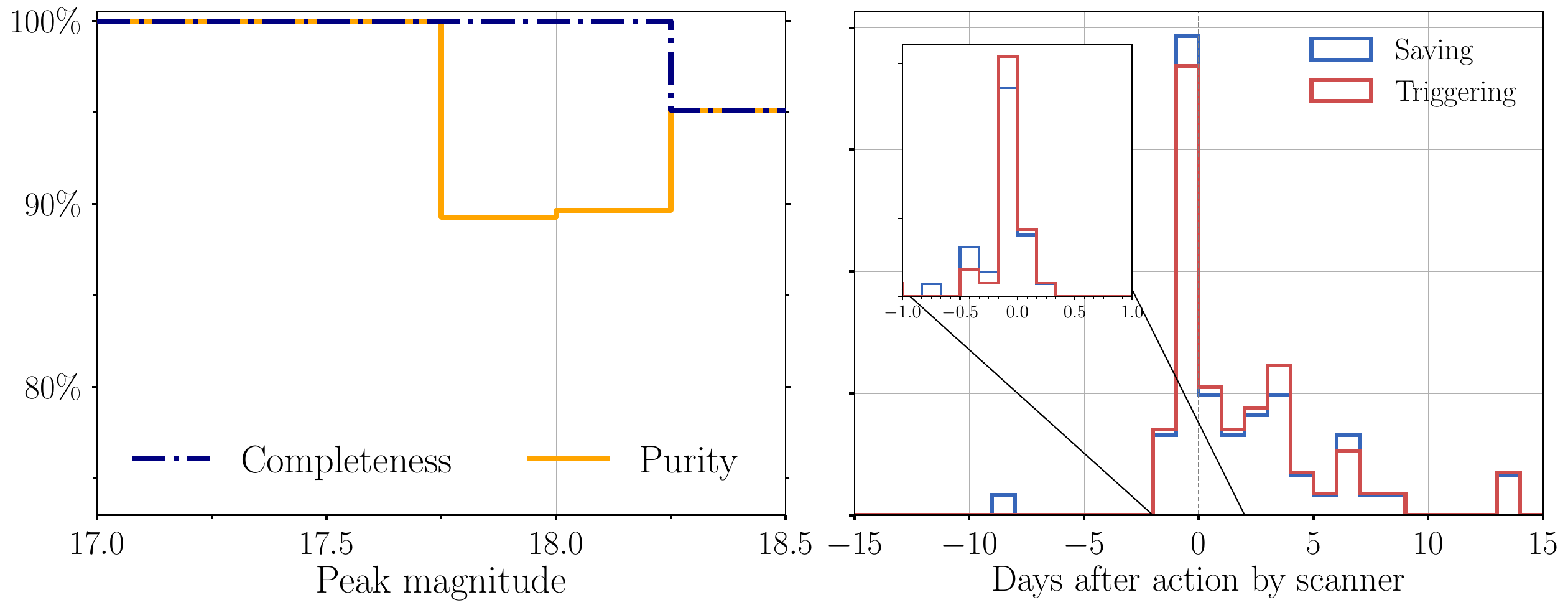}
    \caption{Same as Figure~\ref{fig:test_performance} for a sample of very recent BTS candidates. \textit{Left}: The completeness (dash-dot navy) and purity (solid orange) curves are perfect in most bins. The overall purity of $\texttt{bts\_p1}$ and $\texttt{bts\_p2}$ are \ponepdaypurity~and \ptwopdaypurity~respectively. Contamination unique to $\texttt{bts\_p1}$ is almost entirely real SNe with $m_\mathrm{peak}$ slightly greater than 18.5, acceptable false positives. The relatively low purity in two bins is due to small number statistics. \textit{Right}: Both distributions still peak very near to zero. Likely due to differences in the cuts creating the input samples, $\Delta t_\mathrm{save}$ (\pdaymeddtsave~days) and $\Delta t_\mathrm{trigger}$ (\pdaymeddttrigger~days) are larger than test split equivalents.}
    \label{fig:realtime_performance}
    \end{center}
\end{figure*}

As in Sec.~\ref{sec:test_performance}, any source in \textit{junk} or having only one alert after cleaning is removed when computing policy completeness and purity. In Sec.~\ref{sec:train_data}, the \textit{vars} query received alert thinning down to $N_\mathrm{max}=100$ alerts per source. We emulate this by thinning sources classified on Fritz or manually identified as an AGN, CV, or QSO down to their latest $N_\mathrm{max}=100$ alerts. We run all public alerts passing our Sec.~\ref{sec:train_data} cleaning cuts from all sources through both policies and select only the sources which satisfy $\texttt{bts\_p1}$ or reach their peak magnitude between $2460175.5 < \mathrm{JD} < 2460216.5$. This representatively simulates the actions \BTSbot~would have taken during this time period should it have been fully operational.

Similar to Figure~\ref{fig:test_performance}, Figure~\ref{fig:realtime_performance} shows the completeness, purity, and speed comparison of the production \BTSbot~model on the present-day BTS sample. Both the completeness and purity curve, shown in the left panel, are at or near 100\% in most peak magnitude bins. The overall completeness is \pdaycompleteness. This remains near-perfect and supports \BTSbot's ability to scan without imposing significant selection biases into the BTS sample. Variations in either curve are due to small number statistics; there are 0-3 FPs and FNs in each bin shown. Overall $\texttt{bts\_p1}$ purity is \ponepdaypurity, and overall $\texttt{bts\_p2}$ purity is \ptwopdaypurity. Similar to the test split results, the sources selected by $\texttt{bts\_p1}$ but not $\texttt{bts\_p2}$ are dominated by SNe with $m_\mathrm{peak}$ slightly dimmer than 18.5 mag (19/22 sources): non-problematic FPs because they are typically triggered on by BTS. The completeness and purity estimates of the present-day sample are very similar to their test split analogs, $\leq\pm4\%$ difference in all three metrics, further suggesting that no data shift has affected \BTSbot's performance.

The right panel of Figure~\ref{fig:realtime_performance} compares \BTSbot's speed to act with that of the human scanners on our present-day sample. The medians of $\Delta t_\mathrm{save}$ and $\Delta t_\mathrm{trigger}$ are larger than their test split counterparts: \pdaymeddtsave~days and \pdaymeddttrigger~days respectively. We attribute this increase to two main factors. (i) We have excluded ZTF partnership data from the present-day sample, but scanners typically view data from both the ZTF public and partnership observations when scanning. This frequently provides them with additional detections for BTS candidates, which \BTSbot~is blinded from, aiding in more quickly identify bright transients. (ii) The present-day sample is very small, so the median is volatile to changes. There are just is 65 sources with $\Delta t_\mathrm{save}$ values in the present-day sample, while there were $>250$ sources for the test split. The shapes of the $\Delta t$ distributions, however, is consistent between the present-day and test split analysis. Together, these suggest that the differences observed are not due to a data shift but rather to differences in the sample characteristics and in how the experiments were conducted.

With a sample of BTS candidates contiguous in time, we can now easily compute the median number of saves and triggers performed by \BTSbot~per night. Over the 41 nights in the present-day sample, 184 sources satisfied $\texttt{bts\_p1}$. This is a median of $\sim4.5$ sources per night which would have been saved and sent to SEDM. 

\subsubsection{Analysis of misclassifications in present-day sample}
\label{sec:pday_misclass}

Analogously to Sec.~\ref{sec:test_misclass}, we investigate the types of sources which \BTSbot~misclassifies in the present-day sample. 

Our policies select 35 FPs and 2 FNs. Similarly to the test split results, the majority of FPs (19/35) are real dim transients. The other FPs are again dominated by CVs and AGN but do include other sources like asteroids. One of the FNs (ZTF23aaxtplp) has an extremely bright host galaxy and is projected very near to its nucleus. The other FN (ZTF23abeuope) shows two bright stars overlapping the host galaxy and projected very near to the supernova. It is not certain that these properties caused these sources to receive low scores, but they are clearly very uncommon.
% The other FNs are outlying sources (ZTF23abelseb: an orphan afterglow, and ZTF23aaklqou: SN~2023ixf, a SN in M101), which are rare and atypical owing to their very rapid evolution and their exceptionally bright apparent magnitude respectively. Their atypical nature, however, also makes them two extremely interesting transients. One could design a \BTSbot-like model to automatically identify and trigger follow-up of anomalous transients (see Sec.~\ref{sec:2023ixf} for a discussion of \BTSbot~adaptations). 

As is the case for the test split, the majority of the misclassifications in the present-day sample, dim SNe, have little negative consequence for BTS.
% The remaining misclassifications are mostly CVs, AGN, or QSOs.

%% /------------------------------------------/
%% /--------------- DISCUSSION ---------------/ 
%% /------------------------------------------/
\section{Discussion} 
\label{sec:5}

\subsection{Integration of \BTSbot~into ZTF and the BTS workflow}
\label{sec:integration}

\texttt{BTSbot} has been deployed in \texttt{Kowalski} to enable running in real-time on incoming alert packets from IPAC's alert-producing and brokering system.

\texttt{Kowalski} performs three distinct operations on every alert packet. First, it separates the alert packet from its \texttt{prv\_candidates} field (a 30-day history of detections and non-detections) to concatenate it to the full list of \texttt{prv\_candidates} from all previous alerts with the same \texttt{objectId}. This forms a full light curve for a given ZTF object. The product of this concatenation is then used to compute the custom metadata features \texttt{BTSbot} takes as input (see Sec.~\ref{sec:train_data} for list and definitions). \texttt{Kowalski} then cross-matches every alert with a large number of catalogs such as the NASA Extragalactic Database\footnote{\url{https://ned.ipac.caltech.edu/}} (NED), CLU \citep[Census of the Local Universe;][]{Cook+2019}, Milliquas \citep{Flesch_2019}, and others. 
% Information from the resulting cross-matches is part of the auto-annotations made available to scanners and automated processes. Currently, \texttt{BTSbot} does not use information from these cross-matches but future versions of this model or others may benefit from doing so. 
Lastly, \texttt{Kowalski} runs several ML models: \texttt{braai}, the five ACAI classifiers \citep{Duev+2021}, and \texttt{BTSbot}. The outputs of all models are injected into the alert packet along with their corresponding version numbers allowing alert filters to use this information when identifying candidate transients.
% , which are useful when relying on these models for alert filtering as many groups within ZTF do. 

% Both cross-matches and ML models are entirely configurable from a human-readable configuration file, which allows for a quick iterative process when any updates or additions are required. 
The enriched alert packets are then stored in a non-relational database (MongoDB\footnote{\url{https://www.mongodb.com}}), allowing users to design custom, potentially complex, filters as pure database queries for maximum flexibility. These filters, including the BTS alert filter, run on every alert.

If an alert passes a filter, it will be sent as a candidate to \texttt{Fritz}, a SkyPortal instance that serves as the ZTF collaboration's Marshal. Two additional layers of filtering are applied to allow the automated saving and triggering of any instrument for which SkyPortal has a corresponding application programming interface (API). The first filtering layer assesses if the candidate should be saved, and the second assesses whether an instrument should be triggered for follow-up. Auto-triggering is only run on sources which have passed the auto-saving filter. In the case of \texttt{BTSbot}, two auto-saving and auto-triggering filters implemented, one for each of the policies \texttt{bts\_p1} and \texttt{bts\_p2}.

Additional features are implemented in SkyPortal to ensure that redundant triggers are not sent. A new, automatic follow-up request is prevented if a source already has a spectum, a classification, or if the instrument has already been triggered for that source. Instruments like SEDM which can conduct both spectroscopic and photometric observations get additional rules to define whether the new trigger is redundant with an existing one.
% WIP automatically fetching classifications from TNS to prevent triggers too
Some teams direct scanners to maintain lists of sources that frequently pass their alert filter but are not of interest, e.g. the \textit{junk} set (see Sec.~\ref{sec:train_data}). These can optionally be used to prevent a source from being auto-saved. The payload used for triggering an instrument, which contains the triggering instructions like the priority for SEDM, is set in advance alongside the \texttt{Kowalski} auto-triggering filter, but the priority assigned to a target can be dynamically increased as new alerts are posted. In particular, a source triggered on after passing \texttt{bts\_p1} can have its payload updated to priority 2 if it passes \texttt{bts\_p2} before the trigger is completed.

When auto-saving or auto-triggering actions are taken, comments recording these actions are posted to the relevant source page on \texttt{Fritz}. Beyond bookkeeping, this is crucial to facilitate scanners working alongside \texttt{BTSbot} by displaying \BTSbot's actions in the same interface where manual scanning is performed. Seamless integration with the tools that scanners already rely on enables joint-working that is more efficient and more reliable than either separately. This also allows scanners to modify triggers sent by \texttt{BTSbot}, for example, increasing the trigger's priority or adding photometry to the request. These make real automated triggers safer and more dependable, making the automated aspects of BTS more easily monitored. The results from the SEDM observations and associated data products are uploaded back to Fritz for visualization. 
% The spectral data are visualized in a dedicated plot where scanners can adjust smoothing and overlay spectral lines for a given redshift and expansion velocity. An object's classification, sometimes based on \texttt{SNIascore} \citep{Fremling+2021}, is also displayed using SkyPortal's taxonomy-based classification system.

\texttt{BTSbot}~saved 296 sources to an internal BTS catalog while running in production during October 2023,\footnote{An older version of \BTSbot~(v1.0) was running in production during October 2023. The production model presented here (v1.0.1) corrects for the sources with $m_\mathrm{peak}\leq18.5$ in the \textit{dims} query (see Sec.~\ref{sec:train_data}). Both versions yield nearly identical performance.} and 92.6\% of these were confirmed as extragalactic transients. SN~2023tyk (ZTF23abhvlji) is a Type Ia supernova that was identified and triggered on by \texttt{BTSbot}. The data collected by SEDM was then reduced by \texttt{pySEDM} \citep{Rigault+2019}, the spectrum classified as belonging to a Type Ia SN by \texttt{SNIascore} \citep{Fremling+2021}, and the classification automatically reported to TNS. As detailed by \cite{Rehemtulla_AN23}, SN~2023tyk is the first transient to be fully automatically detected, identified, spectroscopically classified, and publicly reported. As of December 2023, more than a dozen additional Type Ia SNe have been both triggered on by \BTSbot~and spectroscopically classified by \texttt{SNIascore}: SN~2023vcz, SN~2023uxa, SN~2023uti, SN~2023vcx, SN~2023uty, SN~2023vtp, SN~2023vpd, SN~2023vip, SN~2023vwz, SN~2023wts, SN~2023xms, SN~2023xhc, SN~2023xkq. 
% SN~2023uxa (ZTF18aafdigb), SN~2023uty (ZTF23abijmzk), SN~2023wts (ZTF23abnydbs), and SN~2023xhc (ZTF18aaeqjmc).

Operating \BTSbot~and \texttt{SNIascore} alongside each other allows for the full-automation of a significant amount of day-to-day tasks in BTS. About 70\% of bright transients found by BTS are Type Ia supernovae \citep{Fremling+2020}, nearly all of which we expect to be identified by \BTSbot~and 80-90\% of which will be classified and reported to TNS by \texttt{SNIascore}. If no human scanning took place, this BTS workflow could fully-automatically handle more than half of the bright transients in consideration by BTS. In practice, there are additional tasks which remain to be automated or made more efficient. Namely, scanning also involves the retrieving host galaxy redshifts from NED and classifications from TNS. We expect that much of these processes can be automated and planning for doing so is underway. While a comprehensive transition to automated scanning for BTS is unlikely (and not necessarily desirable), these tools mitigate the dependence of BTS operations on the natural fluctuations of scanners' lives. Further, a reallocation of some human effort invested in BTS is possible when many tasks are automated. BTS experts are able to spend more time, e.g., analyzing bulk properties of SN samples or searching for rare or very young events.

\subsection{Comparison with similar models}
\label{sec:other_models}

The ALeRCE\footnote{\url{https://alerce.online/}} \citep[Automatic Learning for the Rapid Classification of Events;][]{Forster+2021} real-time stamp classifier \citep{Carrasco-Davis+2021} and the ACAI \citep[Alert-Classifying Artificial Intelligence;][]{Duev+2021} framework are other image- and extracted feature-based MM-CNNs that perform classification on ZTF alert packets. \BTSbot~is very similar to these models in architecture (indeed, some aspects of \BTSbot~are inspired by them), but they are quite different in application.

The real-time stamp classifier predicts which one of five high-level classes (SN, AGN, VarStar, asteroid, bogus) the source in a ZTF alert belongs to. This is done, primarily, with the goal of automatic and rapid identification of SNe. To this end, the stamp-classifier is trained exclusively on the first alert packet from a given source. While rapid identification of SNe is an interest of ours (see Sec.~\ref{sec:2023ixf}), we instead train \BTSbot~to function on any alert packet from relevant ZTF sources. This grants \BTSbot~the extra utility of being able to identify SNe at any phase of their evolution, albeit with the additional complications associated with doing so. 

The ACAI framework, a union of five independent binary classifiers, predicts which of five phenomenological features (hosted, orphan, nuclear, variable star, bogus) characterizes the source in a ZTF alert packet. These models are trained on any alert from any ZTF source and thus learn a much broader domain than \BTSbot. Narrowing the input domain of our model reduces its broader utility relative to ACAI but unlocks greater performance for our particular task. 

Unlike \BTSbot, neither the stamp classifier nor ACAI learn class definitions that are sensitive to the source's brightness. \BTSbot~must learn to select just a subset of extragalactic transients, those with $m_\mathrm{peak}\,\leq\,18.5\,\mathrm{mag}$, and reject other extragalactic transients and all other sources. While \BTSbot~performs binary classification, effectively marginalizing over all non-bright transient classes, the stamp classifier performs five-class classification and the ACAI models perform five binary-classifications. This requires these other models, the stamp classifier in particular, to learn more discriminatory information between its five classes. This is not necessary for \BTSbot~because our primary interest is automating BTS scanning, for which only bright transients are relevant. Thus, we can justify combining non-bright transients into a single class, simplifying \BTSbot's task.

\subsection{An adaptation of \BTSbot: Automatic, very rapid follow-up}
\label{sec:2023ixf}

The development of \BTSbot~and the design of our policies prioritized completeness of bright transients over other metrics like purity and speed. After operating in production for a few months a different set of policies was implemented, which are focused on increasing purity to operate \BTSbot~with minimal intervention. These priorities are directed by the needs of BTS, but applications of a \BTSbot-like model to other science efforts could prefer to prioritize other metrics. 
% Here, we discuss the strengths of a \BTSbot-like model being used to automatically trigger spectroscopic observations of infant supernovae. 

\BTSbot's architecture is particularly well suited for the automated identification of very young transients. At early times (i.e., the night of the first detection), the source's light curve is uninformative because it is comprised of a very small number of data points. Instead, much of the information available on the new transient is embedded in associated images. In this regime, one would expect that an image-based model, like \BTSbot, would have better access to constraining information than a purely light-curve based model. 

We take SN\,2023ixf as an example illustrative of the additional discovery potential of a model like \texttt{BTSbot}. SN\,2023ixf was a Type II SN in Messier 101. Owing to its proximity, it provided a unique laboratory to study pre-supernova mass loss events of red supergiants, allowed the assembly of a comprehensive time-dependent description of the event, and much more \citep[e.g.;][]{Bostroem+2023, Hiramatsu+2023, Jacobson-Galan+2023, Qin+2023, Zimmerman+2023}. SN\,2023ixf was reported to TNS by Koichi Itagaki at 21:42\,UTC on 19 May 2023 \citep{Itagaki_ixf_TNS}. The earliest published spectrum was collected less than an hour later by \cite{Perley_ixf_TNS}. About 14 hours before the first TNS report, SN\,2023ixf was detected by ZTF, and, just minutes later, this alert packet was assigned a bright transient score of 0.840 by an early version of \texttt{BTSbot}.\footnote{The early version of \BTSbot~referenced here (v0.3) is presented in \citealt{Rehemtulla+2023}.} A variant of \BTSbot~could be trained to isolate such sources. With an alert filter and policy suited for the search of young transients, this could have allowed for a more rapid identification and spectroscopic follow-up of SN\,2023ixf. About half the observing night was remaining for SEDM at the time the first detection would have passed the auto-triggering filters, enough time to obtain a spectrum if triggered at high enough priority. If the spectrum is collected near the end of observing that night at Palomar Observatory ($\sim$12:00\,UTC), this represents a $\sim$10 hour speed-up over the otherwise earliest spectrum obtained. In this example, \texttt{BTSbot} and the associated integration tools presented here allow the probing of early, rapidly-evolving explosion physics typically unavailable to traditional triggering methods.

There are a number of challenges related to this adaptation of \BTSbot. Namely, the BTS alert filter and the \texttt{bts\_p1} and \texttt{bts\_p2} policies presented here would likely be inappropriate or sub-optimal following this new definition of the model's priorities. A thorough exploration of how to best assemble a training set for this goal would also be required.

Automatic spectroscopic follow-up of infant SNe in ZTF data has been successfully implemented before in \texttt{AMPEL} \citep{Nordin+2019}. Target selection was driven by \texttt{SNGuess} \citep{Miranda+2022}, a decision tree-based ML system for identifying SNe. Their automated triggering was designed to observe nearby infant SNe, which was successfully done for a number of sources.

\BTSbot's source code and the production \BTSbot~model are publicly available on GitHub (\url{https://github.com/nabeelre/BTSbot}). This repository includes all codes necessary for assembling \BTSbot's training set, training the model, creating figures visualizing validation or test split performance, and more. It is written specifically with adaptability and flexibility in mind. Additional functionalities, e.g. training models with alternative architectures (see Appendix~\ref{app:unimodal}), are embedded in the scripts with minimal added complexity and no repeated code. This is done to facilitate ease in recycling the \BTSbot~code-base for other applications. One could, for example, quickly explore solving a problem with a simple fully-connected neural network and later advance to a MM-CNN with powerful features like data augmentation and Weights and Biases hyperparameter sweeps integration already built-in. We encourage the use of this resource by the community.

%% /------------------------------------------/
%% /-------------- SUMMARY ---------------/ 
%% /------------------------------------------/
\section{Summary} 
\label{sec:summary}

We have presented \BTSbot, a new multi-modal convolutional neural network to automate scanning for the ZTF Bright Transient Survey. \BTSbot~uses ZTF image cutouts and metadata to produce a bright transient score for an individual ZTF alert packet. It achieves $\sim$95\% accuracy on input alerts and identified 100\% of bright transients in our test split with 93\% purity. Performance compares very closely to that of human scanners in terms of completeness, purity, and speed to act on new transients. \BTSbot~only falls slightly short of scanners in terms of purity of the bright transient sample it produces: $93\%$ vs.\ $\sim97\%$. We also perform an additional analysis with very recent BTS candidates to demonstrate that \BTSbot~is not impacted by a significant data shift.

\BTSbot~has been fully integrated into \texttt{Kowalski} and Fritz, ZTF's first-party alert broker and marshal, and now automatically sends spectroscopic follow-up requests for the new bright transients it identifies. \BTSbot~joins a family of automation tools in the BTS workflow (\texttt{braai}, \texttt{sgscore}, \texttt{pySEDM}, and \texttt{SNIascore}) which aid in running BTS efficiently. These models,  coordinated by Fritz and \texttt{Kowalski}, have enabled the first fully automatic detection, identification, spectroscopic classification, and public reporting of a transient: SN~2023tyk. This milestone represents a boost in efficiency for BTS and an image of what time-domain astronomy could look like during the Rubin-era. It also hints towards the discovery potential of adapting similar technology to other areas of time-domain astronomy.

%% Also note that the akcnowlodgment environment does not support long amounts of text. If you have a lot of people and institutions to acknowledge, do not use this command. Instead, create a new \section{Acknowledgments}.
% \begin{acknowledgments}
\section{Acknowledgments}
A great number of people have contributed to BTS and BTS scanning over the years. We thank the following people who have saved 10 or more sources to internal BTS catalogs on Fritz as of October 2023: Ivan Altunin, Raphael Baer-Way, Pallas A. Beddow, Ofek Bengiat, Joshua S. Bloom, Ola Bochenek, Emma Born, Kate Bostow, Victoria Mei Brendel, Rachel Bruch, Vidhi Chander, Matthew Chu, Elma Chuang, Aishwarya Dahiwale, Asia deGraw, Dmitry Duev, Kingsley Ehrich, Eli Gendreau-Distler, Nachiket Girish, Xander Hall, K-Ryan Hinds, Ido Irani, Cooper Jacobus, Connor Jennings, Joel Johansson, Snehaa Ganesh Kumar, Michael May, William Meynardie, Shaunak Modak, Kishore Patra, Neil Pichay, Sophia Risin, Yashvi Sharma, Gabrielle Stewart, Nora Linn Strotjohann, James Sunseri, Edgar Vidal, Jacob Wise, Abel Yagubyan, Yoomee Zeng, and Erez A. Zimmerman.

We also thank Jakob Nordin for discussions relating to AMPEL.

The material contained in this document is based upon work supported by a National Aeronautics and Space Administration (NASA) grant or cooperative agreement. Any opinions, findings, conclusions, or recommendations expressed in this material are those of the author and do not necessarily reflect the views of NASA. This work was supported through a NASA grant awarded to the Illinois/NASA Space Grant Consortium.

This research was supported in part through the computational resources and staff contributions provided for the Quest high performance computing facility at Northwestern University which is jointly supported by the Office of the Provost, the Office for Research, and Northwestern University Information Technology.

Based on observations obtained with the Samuel Oschin Telescope 48-inch and the 60-inch Telescope at the Palomar Observatory as part of the Zwicky Transient Facility project. ZTF is supported by the National Science Foundation under Grants No. AST-1440341 and AST-2034437 and a collaboration including current partners Caltech, IPAC, the Oskar Klein Center at Stockholm University, the University of Maryland, University of California, Berkeley , the University of Wisconsin at Milwaukee, University of Warwick, Ruhr University, Cornell University, Northwestern University and Drexel University. Operations are conducted by COO, IPAC, and UW.

SED Machine is based upon work supported by the National Science Foundation under Grant No. 1106171

The Gordon and Betty Moore Foundation, through both the Data-Driven Investigator Program and a dedicated grant, provided critical funding for SkyPortal.

N.~Rehemtulla and A.~A.~Miller are partially supported by LBNL Subcontract NO.\ 7707915.

M.~W.~Coughlin acknowledges support from the National Science Foundation with grant numbers PHY-2308862 and PHY-2117997.

SRK thanks the Heising-Simons Foundation for supporting his research.

% \end{acknowledgments}


\vspace{5mm}
\facilities{PO:1.2m, PO:1.5m}
\software{Astropy \citep{astropy:2013, astropy:2018}, corner \citep{corner}, Jupyter \citep{jupyter}, Keras \citep{keras}, Matplotlib \citep{hunter07}, NumPy \citep{numpy}, pandas \citep{pandas1, pandas2}, penquins \citep{penquins}, scikit-learn \citep{sklearn}, SciPy \citep{scipy}, SkyPortal \citep{van_der_Walt+2019, Coughlin_2023}, Tensorflow \citep{tensorflow}, tqdm \citep{tqdm}, and the Weights and Biases platform \citep{wandb}}



\pagebreak
\appendix

\section{Accuracy and loss evolution during training}
\label{app:acc_loss_evo}

\begin{figure}[ht]
    \begin{center}
    \includegraphics[width=0.45\columnwidth]{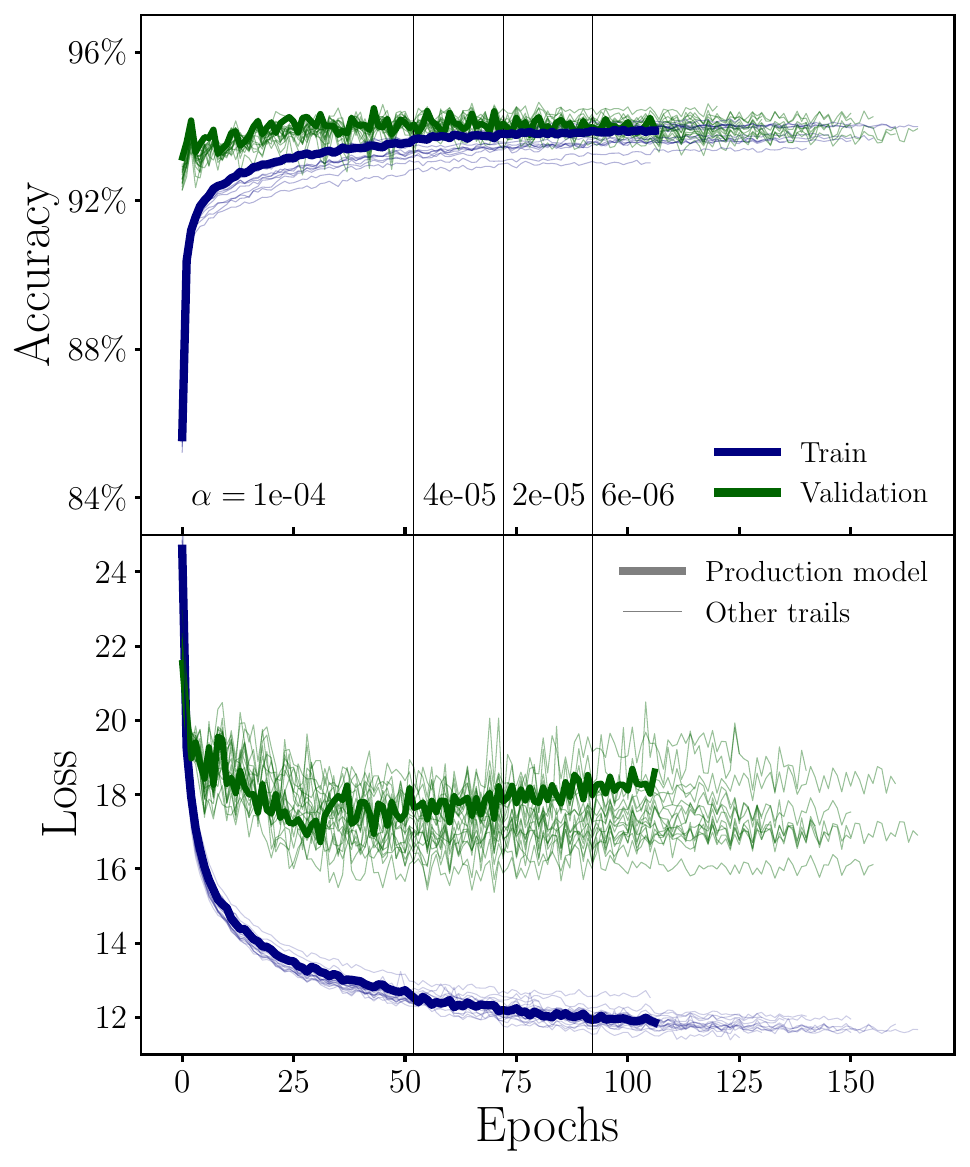}
    \caption{Accuracy (\textit{top}) and loss (\textit{bottom}) for the final \BTSbot~models (production model: bold curve; other trials: narrow curves) of the train (blue) and validation (green) splits during training as a function of epoch. Gray vertical lines indicate epochs at which the learning rate decreased. The discrepancies in these metrics between the train and validation splits cannot be interpreted as overfitting alone because they are partially due to differences in alert thinning and class weighting.}
    \label{fig:acc_and_loss}
    \end{center}
\end{figure}

Figure~\ref{fig:acc_and_loss} shows how the accuracy and loss evolve during the training of the final \BTSbot~models, including the production model. Differences in how the $N_\mathrm{max}$ alert thinning is applied between train and validation (see Sec.~\ref{sec:train_data}) and the use of class weights during training can explain much of the difference between their respective loss curves. Thus, the train loss being significantly lower than validation loss cannot be interpreted as overfitting. There does appear to be overfitting in the production model (bold curves) past epoch $\sim$30, evidenced by the validation loss beginning to increase while the train loss continues to decrease. We avoid this by selecting the model that produced the minimum validation loss during training, rather than the model from the final epoch of training. Figure~\ref{fig:acc_and_loss} marks the epochs where the learning rate decreased due to a plateau in the validation loss as vertical gray lines. The learning rate decreases three times, spanning from $\alpha=10^{-4}$ to $\alpha=6\times10^{-6}$. We also observe that the selected production model is not the model that produced the least validation loss or had the greatest validation accuracy. The production model is selected by considering best performance in policy-based metrics, those most relevant to how \BTSbot~is used.

Training this production model took $\sim$32 hours on an Intel Xeon 6230 CPU, and inference on a single alert packet takes less than a hundredth of a second on a laptop CPU.

\section{Comparison with uni-modal architectures}
\label{app:unimodal}

An exploration of simpler, alternative architectures and a characterization of their performance is necessary to justify \BTSbot's chosen multi-modal architecture. We present alert- and policy-based performance metrics produced by two uni-modal model architectures: a uni-modal convolutional neural network (UM-CNN) and a fully-connected neural network (NN). These networks perform inference using only images and only metadata respectively. The training data, image preprocessing, and feature extraction are, where relevant, performed identically for all three architectures. The ordering and types of layers of the uni-modal architectures are identical to those of the multi-modal architecture with the other branch entirely removed. The hyperparameters of the layers, the Adam optimizer, and the value of $N_\mathrm{max}$ are searched over in a Bayesian hyperparameter sweep similarly to the sweeps executed for the MM-CNN. 

Table~\ref{tab:unimodal_archs} shows the optimal layer hyperparameters for either architecture. Compared to the MM-CNN's convolutional branch, the UM-CNN has smaller convolutional kernels and a much larger dense layer following the flattening. The UM-CNN's first two convolutional layers have more filters than the final two convolutional layers, whereas the pattern is reversed for the MM-CNN. We also find that the optimized UM-CNN has $\textrm{batch size}=32$, and Adam parameters $\beta_1=\beta_2=0.999$. Compared to the MM-CNN's metadata branch, the optimized NN has half as many neurons in the first two dense layers but eight times as many in the third layer. We also find that the NN performs best with $\textrm{batch size}=128$, Adam parameters $\beta_1=\beta_2=0.999$, and $N_\mathrm{max}=30$. The hyperparameter values which are not explicitly mentioned can be assumed to be identical to that of the MM-CNN.

\begin{deluxetable}{cc|cc}
\tablecaption{Architectures of uni-modal \BTSbot~alternatives \label{tab:unimodal_archs}}
\tablehead{
    \multicolumn{2}{c}{Uni-modal convolutional neural network} & \multicolumn{2}{c}{Fully-connected neural network}
}
\startdata
    Layer type & Layer parameters & Layer type & Layer parameters \\
    \hline
    2D Convolution  & 64 filters, $3\times3$~kernel     &     Batch normalization & - \\
    2D Convolution  & 64 filters, $3\times3$~kernel     &     Dense               & 64 units \\
    Max pooling     & $2\times2$~kernel                 &     Dropout             & 0.40 \\
    Dropout         & 0.45                              &     Dense               & 64 units\\
    2D Convolution  & 16 filters, $3\times3$~kernel     &     Dense               & 64 units \\
    2D Convolution  & 16 filters, $3\times3$~kernel     &     Dropout             & 0.70 \\
    Max pooling     & $4\times4$~kernel                 &     Dense               & 1 unit \\
    Dropout         & 0.65                              &     \\
    Flattening      & -                                 &     \\
    Dense           & 128 units                         &     \\
    Dropout         & 0.45                              &     \\
    Dense           & 1 unit                            &     \\
    \hline
\enddata
\tablecomments{Layer parameters of alternative \BTSbot~architectures: uni-modal convolutional neural network (left two columns) and fully-connected neural network (right two columns). Optimal hyperparameters are determined by large hyperparameter sweeps.}
\end{deluxetable}

\begin{deluxetable}{c|cc|ccccc}
\tablecaption{Performance metrics of uni-modal \BTSbot~alternatives \label{tab:arch_comparison}}
\tablehead{
    \multirow{2}{*}{Architecture type} & \multicolumn{2}{c|}{Alert-based metrics} & \multicolumn{5}{c}{Policy-based metrics} \\
    & \colhead{Accuracy} & ROCAUC & Completeness & \colhead{\texttt{bts\_p1} purity} & \colhead{\texttt{bts\_p2} purity} & \colhead{$\Delta t_\mathrm{save}$} & \colhead{$\Delta t_\mathrm{trigger}$}
}
\startdata
    NN & \textbf{95.5\%} & \textbf{0.988} & 99.8\% & 82.8\% & 91.2\% & -0.0400 days & \textbf{-0.0200 days} \\
    UM-CNN & 82.4\% & 0.902 & 94.3\% & 69.5\% & 92.5\% & \textbf{-0.0485 days} & -0.0174 days \\
    MM-CNN & \testacc & \testrocauc & \textbf{\testcompleteness} & \textbf{\ponetestpurity} & \textbf{\ptwotestpurity} & \testmeddtsave~days & \testmeddttrigger~days
\enddata
\tablecomments{Comparison of performance metrics on test split data across three different model architectures: a fully-connected neural network (NN), a uni-modal convolutional neural network (UM-CNN), and a multi-modal convolutional neural network (MM-CNN). Each network is the highest performing in at least one metric. The UM-CNN is less complete than the MM-CNN and both the NN and the UM-CNN fall short of the MM-CNN in purity. Overall, the MM-CNN is the best performing architecture.}
\end{deluxetable}

Table~\ref{tab:arch_comparison} compares test split performance metrics from each of the three architectures. The NN slightly outperforms the MM-CNN in the alert-based metrics while the UM-CNN falls well short of the others. The UM-CNN presented here uses the full-size image cutouts, but Appendix~\ref{app:cropped} shows that slightly increased performance can be realized through the reduction of input image's size. In policy-based metrics, the MM-CNN's marginal advantage in completeness over the NN is due to having just a single fewer FN. The MM-CNN demonstrates a larger advantage in purity, however. Its advantage is $\sim0.5-2\%$ over the others in \texttt{bts\_p2} purity; even small boosts in purity are valuable when applied over the large samples and baselines in consideration by BTS. Each model performs very similarly in the speed metrics, although the NN and the UM-CNN marginally outperform the others in $\Delta t_\mathrm{trigger}$ and $\Delta t_\mathrm{save}$ respectively.

Each architecture outperforms the others in some metrics, but the MM-CNN delivers the best overall performance. It delivers the highest completeness and purity while sacrificing well less than an hour in $\Delta t_\mathrm{save/trigger}$. 

Despite having worse purity than the MM-CNN, there is still valuable utility in the NN because we find that it trains $\sim60$ times faster than the MM-CNN. This discrepancy in training time is partially due to the NN adopting $N_\mathrm{max}=30$ instead of $N_\mathrm{max}=100$ as the MM-CNN does, and could be also decreased by training the MM-CNN on a GPU. These factors aside, the NN would very likely remain many factors quicker to train, and it is thus better suited for experimenting with adapting \BTSbot~to alternative use cases. In developing adaptations of \BTSbot, being able to very quickly design and execute experiments will be key. Excluding images may also lend advantages in the ease of transferring \BTSbot~adaptations to other surveys, for example, LSST or the upcoming La Silla Schmidt Southern Survey (LS4\footnote{\url{https://www.snowmass21.org/docs/files/summaries/CF/SNOWMASS21-CF6_CF4_Peter_Nugent-171.pdf}}). Although effort has been made to develop survey-agnostic CNNs \citep[e.g.,][]{Cabrera-Vives+2023}, \BTSbot~is not designed or expected to perform consistently on image data from other surveys. Instead, one could more easily design a NN which exclusively uses survey-agnostic features, increasing its potential impact by allowing it to be applied more widely.

\section{Convolutional neural networks with cropped image cutouts}
\label{app:cropped}

The properties of the image cutouts produced by large wide-field surveys are critical to a number of the survey's functions. While the pixel scale, typically measured in arcseconds (") per pixel, is a fixed property of the telescope's optics, many choices can be made in software that determine the nature of the cutouts sent to alert brokers. One must choose the angular and pixel size of the cutouts in light of the pixel scale and whether or not to decrease resolution by binning the pixels. A tension arises because scanners generally prefer having more information, i.e. cutouts with higher resolution and greater angular size, but the surveys producing and disseminating the alerts prefer minimal network bandwidth costs, i.e. lower resolution and smaller angular size. Smaller cutouts with less information may compromise the performance of these key ML models, however, possibly reducing the scientific potential of the entire survey. Only a small number of studies developing CNNs for large wide-field surveys comment on performance over a range of cutout sizes \citep[e.g.;][]{Killestein+2021, Carrasco-Davis+2021, Reyes-Jainaga+2023}. It remains to be seen, then, what the minimum cutout size is to maintain acceptable performance with current CNNs. Adding further complexity, CNNs performing different tasks will be impacted differently by reduced cutout sizes, and MM-CNNs may be more resilient to reduced cutout sizes due to the availability of metadata information. Multi-scale cutouts, where resolution progressively decreases moving away from the cutout center, are a compelling alternative because they dramatically increase the available field-of-view without an increase to the data volume of the cutouts \citep{Reyes-Jainaga+2023}. Further study of (MM-)CNN performance over a range of cutout sizes is necessary to better characterize their correlation.

\begin{figure}[ht]
    \begin{center}
    \includegraphics[width=0.5\columnwidth]{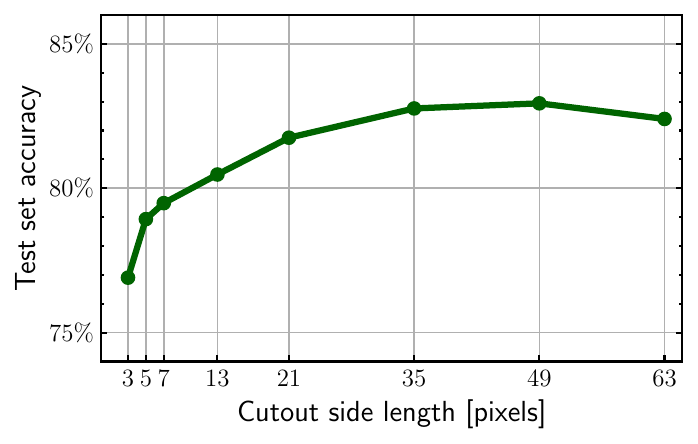}
    \caption{Test set accuracy of \BTSbot~UM-CNNs as a function of input image cutout size. Very small cutouts ($3 \times 3$ to $13 \times 13$ pixels) clearly under-perform relative to models with larger cutouts ($35 \times 35$ to $63 \times 63$ pixels). The highest performing model is notably not that which uses the full-size images but rather the model which uses $49 \times 49$ pixel cutouts. Cropping cutouts could allow for the significant decrease of a survey's data rate and possibly network bandwidth costs, in this case, with no decrease in the performance of ML-based transient detection tools.}
    % \nabeel{This may be too difficult if the sweeps are not ``complete" but don't you also have some sense of the variance based on the results from each of the individual sweeps/runs?}
    \label{fig:cropped_performance}
    \end{center}
\end{figure}

To explore this important issue, we train multiple UM-CNN versions of \BTSbot~to identify which cutout size produces the most accurate model. We create new training sets from \BTSbot's original training set by keeping only the innermost $N_\mathrm{pix} \times N_\mathrm{pix}$ pixels of each cutout and renormalizing them individually. We train $\sim$30-60 trials of \BTSbot~with a given value of $N_\mathrm{pix}$ (3, 5, 7, 13, 21, 35, or 49) in Bayesian hyperparameter sweeps similar to Section~\ref{sec:train_and_opt} and Appendix~\ref{app:unimodal}. These trials are unlikely sufficient to yield optimal hyperparameters, but they do give a representative view of the performance for each value of $N_\mathrm{pix}$. For models with $N_\mathrm{pix}\leq13$, we remove the pooling layers entirely to maintain the image's size through the hidden layers.

Figure~\ref{fig:cropped_performance} presents the highest performing models judged by accuracy on the test set. We find that the $N_\mathrm{pix}=49$ UM-CNN yields the best performance, and that both the $N_\mathrm{pix}=49$ and $N_\mathrm{pix}=35$ UM-CNNs marginally outperform the UM-CNN with uncropped cutouts ($N_\mathrm{pix}=63$). It is not unambiguously clear what the reason for this is from this simple experiment alone, however, these results suggest that information $\sim0.5'$ away from the source in question tends to be irrelevant or noisy when performing our task. We notably do not observe the same trend for the MM-CNN, where the $N_\mathrm{pix}=63$ model is the highest performing. \cite{Carrasco-Davis+2021} find that the ALeRCE stamp classifier (a MM-CNN), performs best with $N_\mathrm{pix}=21$ cutouts. Together, this indicates that more work is necessary to better understand the optimal cutout size for upcoming surveys.

The $N_\mathrm{pix}=35$ model roughly matches the $N_\mathrm{pix}=63$ model in accuracy but has $35^2 / 63^2 \approx 30\%$ the number of pixels. In this case, shrinking cutouts dramatically quickens \BTSbot's training without compromising performance. Shrinking cutouts survey-wide is an option to reduce a survey's network bandwidth costs significantly as image cutouts typically comprise a large fraction of the bytes in an alert packet. Although $35 \times 35$ cutouts are appropriate for \BTSbot, such small cutouts may reduce scanners' ability to distinguish sources of different types. Experiments of this sort are most relevant to the alert stream design of upcoming surveys like LSST and LS4. LSST's planned alert packet cutout size is at least $30 \times 30$ pixels\footnote{\url{https://dmtn-102.lsst.io}} at a pixel scale of 0.2"/pixel \citep{Ivezic+2019}, thus spanning 6" on a side. The $N_\mathrm{pix}=5$ and $N_\mathrm{pix}=7$ models are nearest to this in terms of field-of-view, however, ZTF's much larger pixel scale (1"/pixel) makes this an unreasonable comparison. Further study is required to assess how small cutouts can be without compromising, e.g., real/bogus performance at LSST-like resolutions. LS4 will have a pixel scale very similar to that of ZTF, but, at the time of writing, the alert packet cutout size is undetermined (R. Knop, private communications). These results provide the preliminary guidance that LS4 cutouts should be no smaller than $21 \times 21$ in order to maintain the performance of \BTSbot-like UM-CNN models.

% comment on random guessing accuracy = 65\%

% with small crops we could rotate by an arbitrary angle and not have to make up pixel values

% For example, a very faint alert that shows next to no features in its images can be identified as an AGN by a MM-CNN because AGN often have a large number of previous detections at their location (represented by \texttt{ndethist}), information a uni-modal CNN could not have.


\bibliography{tda_refs}{}
\bibliographystyle{aasjournal}



\end{document}